\documentclass{pasj00}
\draft

\begin{document}
\SetRunningHead{Author(s) in page-head}{Running Head}
\Received{2010/01/22}
\Accepted{2010/05/24}

\title{Oscillations in the G-type Giants }

\author{Hiroyasu \textsc{Ando}$^{1,2,3}$, Yusuke \textsc{Tsuboi}$^{3}$,  Eiji \textsc{Kambe}$^{4}$, and Bun'ei \textsc{Sato}$^{5}$}%
\affil{$^{1}$National Astronomical Observatory of Japan, National Institutes of Natural Sciences, 2-21-1,Osawa, Mitaka, Tokyo, 181-8588}
\email{ando.hys@nao.ac.jp, spfc6ec9@diary.ocn.ne.jp}
\affil{$^{2}$Department of Astronomy, Graduate University for Advanced Studies, Shonan Village, Hayama, Kanagawa, 240-0193, Japan}
\affil{$^{3}$Department of Astronomy, Graduate School of Science, The University of Tokyo, 7-3-1 Hongo, Bunkyo-ku, Tokyo, 113-0033, Japan}
\affil{$^{4}$Okayama Astrophysical Observatory, National Astronomical Observatory of Japan, National Institutes of Natural Sciences, 3037-5, Honjo, Kamogata, Asakuchi,Okayama 719-0232, Japan}\email{kambe@oao.nao.ac.jp}
\affil{$^{5}$Tokyo Institute of Technology, 2-12-1, Ookayama, Meguro-ku, Tokyo, 152-8550, Japan}\email{sato.b.aa@m.titech.ac.jp}

%


\maketitle

\begin{abstract}
The precise radial-velocity measurements of 4 G-type giants, 11Com, $\zeta$ Hya, $\epsilon$ Tau, and $\eta$ Her were carried out.  The short-term variations with amplitudes, 1-7m/s and periods, 3-10 hours were detected.  A period analysis shows that the individual power distribution is in a Gaussian shape and their peak frequencies ($\nu_{max}$) are in a good agreement with the prediction by  the scaling law.  With using a pre-whitening procedure, significant frequency peaks more than 3 $\sigma$ are extracted for these giants.  From these peaks, we determined the large frequency separation by constructing highest peak distribution of collapsed power spectrum, which is also in good agreement with what the scaling law for the large separation predicts.  Echelle diagrams of oscillation frequency were created based on the extracted large separations, which is very useful to clarify the properties of oscillation modes.  In these echelle diagrams, odd-even mode sequences are clearly seen.  Therefore, it is certain that in these G-type giants, non-radial modes are detected in addition to radial mode.  As a consequence, these properties of oscillation modes are shown to follow what  \cite{D-2}(2001) and \cite{D-5}(2009) theoretically predicted. Damping times for these giants were estimated with the same method as that developed by \cite{S-3}(2004).  The relation of Q value (ratio of damping time to period) to the period was discussed by adding the data of the other stars ranging from dwarfs to giants. 
\end{abstract}

\section{Introduction}
The solar-like oscillations are now well known to exist ubiquitously in the late type stars from dwarfs to giants owing to  intensive observations by many groups in ground-based spectroscopy and space-based photometry  (see  \cite{B-2}2007).   An asteroseismic approach by comparing observed mode frequencies  with theoretical model frequencies has been conducted extensively to investigate stellar evolutionary stages, masses, and their internal structures (ex., \cite{M-1}(2005) for $\alpha$ Cen A,B, and \cite{K-1}(2008) for $\epsilon$ Oph).   Nowadays, the very accurate fitting of the observed frequencies to the model frequencies has allowed us to obtain better understandings of the stellar evolution and the structure (see, \cite{K-1}2008).  Even so, there are still debates on the properties of oscillation modes (radial or non-radial, and long or short lifetime of modes) in giants (see \cite{D-2}2001 and \cite{D-1}2009).   It is certain that to obtain more comprehensive and consistent understandings of the stellar evolution and the structure, much higher precision observations, and non-interrupted and long-term observations are needed for as many giants as possible. \\
\indent In ground-based spectroscopic studies of radial-velocity variations, the detection of the solar-like oscillations has been devoted mainly to dwarfs and subgiants, since their short periodicities may give a guarantee for sufficient coverage of the oscillation cycles in a  plausible observation span allowed for  ground-based telescopes.  For giant stars, to our knowledge three G-type giants ($\xi$ Hya, $\zeta$ Hya, and $\epsilon$ Oph) and eight K-type giants ($\eta$ Ser, $ \beta$ Gem, $\alpha$ Ari, $\alpha$ UMa, $\beta$ Oph, $\iota$ Dra, $\alpha$ Boo, and $\gamma$ Dra) were observed.  Their amplitudes are relatively large enough to be detected, but their cycle coverages are not enough to examine their periodgram in detail. \\ 
\indent  On the other hand, in space-based photometric studies, long-term and continuous observations for many red giants have recently been performed by dedicated satellites, for instance, CoRoT(see \cite{H-1}2010), and Kepler (see \cite{G-5}2010).  They discussed the characteristics of solar-like oscillations concerning their respective first results of several hundreds of red giants.  \cite{D-1}(2009)  showed  clear odd-even mode sequences in echelle diagrams of the oscillation frequencies in some red giants, and pointed out that non-radial oscillations are clearly detected in addition to radial modes.  They also suggested that there are some giants showing the complex power spectra, and the straightforward explanations are not appropriate. For 50 low-luminous($L \le 30 L_{\odot}$) red giants, \cite{B-5}(2010) showed  clear odd-even mode sequences to point out that modes with $l=1$ having a slightly large scatter in their frequency distribution behave like a mixed-mode and some modes even with $l=3$ are identified, which demonstrates highly accurate photometry of Kepler satellite.  \\
\indent \cite{D-2}(2001)  discussed the differences in the non-radial mode properties in giants depending on their temperature and luminosity. \cite{D-5}(2009) further developed their analysis with more elaborate methods of evaluating damping and excitation of oscillations along a red giant branch. The key physical point is  the depth of the convection zone.  In a high red giant branch (higher luminosity), the surface convection zone that separates p-mode(surface) and $\it{g}$-mode propagation zones (core)  is so deep that the interaction of the p-mode and the $\it{g}$-mode becomes very weak, and a non-radial  p-mode called as STE (Strongly Trapped in the Envelope) mode is excited with an observationally large enough amplitude.  A regular equally-spacing pattern is clearly seen in the power spectrum.  On the other hand, in an intermediate red giant branch and/or at its bottom, the surface convection zone is shallower, so that the interaction of the p-mode in the envelope and the $\it{g}$-mode in the core, particularly with degree $\it{l} =1$, becomes strong, and  such a mixed-mode (avoided crossing) with a somewhat shifted frequency from the intrinsic position for isolated p-mode potential has a smaller amplitude.  \cite{D-5}(2009) pointed out that it is very important to pay attention to the height of a mode instead of its amplitude, since the heights are directly related to the observational peaks in the power spectrum.  They showed that in an intermediate red giant branch, the radial modes and non-radial modes trapped in the envelope are observable with similar heights in the power spectrum.  However, trapping is not perfect for modes with $\it{l} =1$, and thus a small group of $\it{l} =1$ modes is detectable around each local maximum in the power spectrum corresponding to an envelope trapped mode.  At the bottom of a red giant branch, the heights in the power spectrum for radial and non-radial modes are also similar, which would be detected in the power spectrum.  However, the interaction between the p-mode and $\it{g}$-mode is so strong that many avoided crossings occur.  The resulting frequency pattern in the power spectrum is very complex.        \\
\indent Despite observational difficulty in the ground-based spectroscopic studies, we tried radial-velocity monitoring of 4 G-type giants (11Com, $\zeta$ Hya, $\epsilon$ Tau, and $\eta$ Her) to investigate the properties of the solar-lke oscillations.  In this paper, we report on the detection of the solar-like oscillations in these targets.  Section 2 provides observation and data reduction, and a period analysis is given in section 3.  Section 4 is dedicated to conclusions. \\
\section{Observations and Radial-Velocity Analysis}
We selected 4 G-type giants among a target list of the exoplanet search program carried out at Okayama Astrophysical Observatory (\cite{S-1}2002).   Their locations in H-R diagram are shown in figure 1 together with stars that were reported so far to show evidence of solar-like oscillations. As can be seen from this figure, 4 giants are distributed from less-luminous ($L < 100 L_{\odot}$) G giant to luminous ($L > 100 L_{\odot}$) G giant, respectively.  This sample is appropriate to investigate the oscillation properties in giants, as mentioned in the introduction.  \\
\indent For these stars, a spectroscopic observation with an iodine absorption cell (I$_2$ cell) was carried out from 2006  January to 2009 March, using HIDES (HIgh Dispersion Echelle Spectrograph) at the coud$\acute{\rm e}$ focus of the $74^{\texttt{"}}$ reflector at Okayama Astrophysical Observatory (OAO). For radial-velocity measurement, the wavelength region was set to cover 5000-6100A and the slit width was set to 250 $\mu$ m, corresponding to a spectral resolution of $R \sim 59,000$ (about 3.8 pixels sampling).  We  obtained spectra typically with S/N$\simeq 250 ~\rm{pix} ^{-1}$ . To make a template spectrum  for each target, we took more than 10 spectra without an I$_2$ cell using a 100 $\mu$m slit width ($R \sim 112,000$). The typical S/N of the resulting templates  was about $750 ~\rm{pix} ^{-1}$. The journal of the observations for these targets  is given in table 1. Two stars, 11 Com and $\zeta$ Hya among our sample were observed twice in different seasons to check whether their characteristics of the radial-velocity variations may change over a long time span. \\
\indent A data reduction of echelle spectra (i.e., bias subtraction, flat-fielding, scattered-light subtraction, and spectrum extraction) was performed for these stars using the IRAF software package in the standard manner. Concerning the wavelength calibration and  subsequent radial-velocity measurements, the same procedures as those in \cite{A-1}(2008) were used.  All reduced velocities were corrected for the Earth's motion against the barycenter. \\
\indent Here we show the time variations in the radial velocity for individual star. \\
$11 \ Com$ \\
Figure 2a shows that the time coverage of the observation in 2008 was not bad during the observed night, but the data were actually limited only for four nights due to the weather conditions. On the other hand, the observations in 2009, as shown in figure 2b, had  a relatively good time coverage.  From these, the variations on a time scale of about 10 hr can be conspicuously seen. In these figures, the upper panel indicates the  internal error for each observational point. \\
$\zeta \ Hya $\\
The variations in 2006 and in 2008 are shown in figure 3a and 3b, respectively. The observation in 2006 was quite good.  However, the time coverage for each night in 2008 was a little poor due to bad weather.  Even so, periodic variations of a time scale ranging from 6 to 10 hr are apparently seen. \\
\\
$\epsilon  \ Tau$  \\
In figure 4, the variations on a time scale ranging from  4 to 5 hr can be clearly seen.  \\
$\eta \ Her$ \\
Figure 5 shows the variations of the time scale ranging from 3 to 4 hr, although the time coverage was not satisfactory.  
\\
\indent Long-term variations with a time scale of  several days are apparently seen for these stars.    Such variations were also discovered in a campaign observation of Procyon, conducted in 2006-2007 (see \cite{A-2}2008), in which its origin has been considered to be related to rotational modulation.  However,  similar variations of our targets are not  discussed here.  In a subsequent analysis, any long-term variations were removed using a pre-whitening technique. \\
\indent The stellar parameters for these giants are given in table 2.   The surface gravities ($\rm{log} \hspace{1pt} \it{g}$) of $\zeta$ Hya recently given by \cite{S-7}(2005) and \cite{T-1}(2008) are 2.65 and 2.30, respectively. The discrepancy is a little large.  Considering the value of $\rm{log} \hspace{1pt} \it{g}$= 2.5 given by \cite{L-2}(1995),we here adopt the average value (2.48) of the two as  $\rm{log} \hspace{1pt} \it{g}$ of $\zeta $ Hya.  The radii were reduced from the revised Hipparcos parallax (\cite{L-1}2007), and the apparent diameters were measured by interferometry; $\zeta$ Hya and $\eta$ Her (\cite{D-3} 2005), and $\epsilon$ Tau (\cite{M-2} 2003).  However, the interferometric apparent diameter for 11 Com is not available, to our knowledge.

\section{Period Analysis}
\subsection{Power distribution}
To investigate the frequency distribution of the power for the short-term variations of 4 G-type giants, we made a discrete Fourier transform (\cite{S-2}1982) for the time series of these giants.  The power spectra for 11Com, $\zeta$ Hya, $\epsilon$ Tau, and $\eta$ Her are shown in figure 6a,b, figure 7a,b, figure 8, and figure 9, respectively.   The power excess above the noise level estimated at the higher frequencies (typically 150-200 $\mu$Hz) is clearly seen in each spectrum.  It is very interesting to note that in both cases of 11 Com and $\zeta$ Hya, the shape of the power distributions taken during different seasons and under different observational conditions are quite similar, even though the amplitudes of individual peaks changed.   This means that the basic excitation mechanism did not change.  \\
\indent We performed smoothing of the power spectrum by convolving with a Gaussian having a FWHM of 4-times  the expected large separation for each giant.  A resulting smoothed power distribution gives a Gaussian-like power distribution (dotted line), as shown in these spectra.   If we regard the peak frequency as $\nu_{max}$, proposed by \cite{K-2}(1995), these peak frequencies are, as indicated in table 3, in a  good agreement with the predicted values from the scaling law for $\nu_{max}$, such as $\nu_{max} = 3050 ( g/g_{\odot} ) ( Te/Te_{\odot})^{- 1/2}$  ( $\mu$Hz ).   \\
\indent From the stellar parameters given in table 2, we can calculate the acoustic cut-off frequency at the temperature minimum, above which oscillations are propagating through the photosphere to the outer region of a  star.  Here, we assume a scaling law for the acoustic cut-off frequency ($\nu_c$) to the Sun, such as $\nu_c = 5400(g/g_{\odot})(Te/Te_{\odot})^{-1/2}$ ($\mu$Hz).  In figures 6-9, such frequencies are indicated by vertical dashed lines, respectively.  It is certain that in their power spectra, there is weak, but significant, power excess beyond these acoustic cut-off frequencies.  It should be reminded that in the solar 5-minute oscillations,  a weak power excess (less than 1/10 of peak power) is observed beyond its acoustic cut-off frequency($\nu_{ac}$= 5.4mHz $< \nu < $7.5mHz), as shown by \cite{G-3}(1998).  However, it is difficult to explain these significant power excess beyond the acoustic cut-off frequencies by the solar analog.  This means that a more realistic atmosphere having a higher acoustic cut-off should be considered in these G-type giants.    \\
\subsection{Large separation}
Following the procedure done by \cite{A-1}(2008) , a Fourier transform and the pre-whitening procedure were applied repeatedly to extract significant peaks until the amplitude of a newly searched peak became smaller than  the noise level ($\sigma$)  in the velocity amplitude (not in the power) at  higher frequencies.   We also carefully checked if the comparable higher adjacent peaks with a spacing of about 11.57$\mu$Hz coming from one-day aliasing could be seen.  If so, we tried peak subtraction for every peak, and compared its remaining standard deviation  to seek the smallest one.   Of course, this process cannot completely exclude false peaks, but is effective to reduce false peaks.  From such a detailed analysis, we picked up peaks for 4 G-type giants whose amplitudes were greater than 3$\sigma$.  These are summarized in table 4 to 7 .  \\
\indent  To derive  large frequency separations ($\triangle \nu$) for these 4 giants, we calculated the auto-correletion and the comb response function from these power spectra, following  previous studies.  However, we found no conspicuous peaks, except for those peaks related to one-day aliasing.  If some peaks are missing in the spectrum due to mode coupling or mode damping, these correlation methods do not work effectively.  In fact, the situation of missing modes in the solar-like oscillations is the case for giants.   \\
\indent Instead, we follow here an idea proposed by \cite{C-2}(2007), and applied practically by \cite{T-4}(2009).  In their method, the highest peak in the collapsed power spectrum is searched for a range of values of a large separation.  The collapsed power spectrum for a given value of $\triangle \nu$ is calculated by dividing the power spectrum into segments of length $\triangle \nu$ and summing these (see \cite{G-2}(1980) for the Sun).      In a subsequent analysis, we  modify the original procedure as follow. Instead of using the observed power spectrum, we use the artificial power spectrum so as to avoid  heavy disturbances from one-day aliasing and noise.   The artificial power spectrum is constructed  in such a way that  the extracted peaks given in table 4-7 for each giant have Gaussian PSFs with the same amplitude and width corresponding to observational span,  distributed over the frequency range (0-200 $\mu$Hz).  If the extracted power, itself, is used as the peak amplitude, the regular pattern suffers from a severe deformation by a strong peak.  Our aim is to find a regular pattern.  If the extracted frequencies are described by a well-known asymptotic formula, such as $\nu = \triangle \nu (n + l/2 + \epsilon ) + X_0$($n$: radial order, $l$: degree),   the highest peak of the collapsed power spectrum should be found at a location of  1/2$\triangle \nu$  and, accordingly, its corresponding local peak  would be found at a location of $\triangle \nu$.  If the modes  are only radial pulsations ($l=0$), the only one sequence at both of 1/2$\triangle \nu$ and $\triangle \nu$ should  emerge. \\
\indent Following this procedure, we constructed the peak distribution of the collapsed power spectrum over a range of values of the large separation (1-9 $\mu$ Hz) for each giant.  As an example, the peak distribution of the collapsed power spectrum for 11Com is shown in figure 10, where only data in 2009 were used due to the lower noise and better data sampling.  This figure clearly shows the highest peak at 1.44 $\nu$Hz (1/2$\triangle \nu$),  and also a local peak is found at 2.88 $\nu$Hz ($\triangle \nu$).  As a consequence, we derive  large separations for 4 giants, which are given in the table 3.   \\
\indent Based on  the extracted large separations, the echelle diagrams of the oscillation frequencies for 4 giants were constructed, as shown in figures 11-14, respectively, to illustrate the properties of the oscillation modes.  In the power spectrum, the frequency axis is divided into segments of length $\triangle \nu$, and the segments are stacked vertically like an optical echelle spectrum.   For a comparison, in the table 3, we present the large separation calculated from the scaling law for a large separation proposed by \cite{K-2}(1995), such as $\triangle \nu  =  134.9( g/g_{\odot}) ^{1/2} (R/R_{\odot}) ^{-1/2} $   ( $\mu$Hz ).  Obviously, the observational large separations for our giants are in  good agreement with the predicted ones from the scaling law.  \\
\indent It should now be pointed out that in their echelle diagrams, odd-even mode sequences are clearly seen, although one sequence has numerous modes and another sequence has only one mode in the less luminous giants, $\epsilon$ Tau and $\eta$ Her. This fact indicates that non-radial modes are detected in addition to radial modes in G-type giants.   The radial mode sequence ($l=0$) with higher order (large $n$) is well aligned vertically in an echelle diagram, since the radial mode does not interact with $g$-mode. On the other hand, the non-radial p-mode can interact with the $g$-mode to become a mixed-mode, which suffers  from a significant frequency shift.  In the dipole mode ($l=1$) sequence, a mixed-mode with a significant frequency shift is particularly conspicuous, manifesting a large scatter in the frequency distribution. From these characteristics, we can somehow distinguish which is which.  However, we should postpone any conclusion on which sequence corresponds to the odd modes ($l=1, 3,....$) and the other to even modes ($l=0,2,....$) until a detailed comparison of the observation with the theoretical model can be made.     \\
\indent  It is now certain that our findings just mentioned above lead us to mention what \cite{D-2}(2001) and \cite{D-5}(2009) pointed out.  In 11 Com, an odd-even sequence pattern in the echelle diagram is clearly seen, which is very similar to  case C discussed by \cite{D-5}(2009).  In high red-giant branch, the convection zone is  so deep that the amplitude of STE mode with $l=1$ becomes large enough to be observable in addition to radial pulsation modes. In a less-luminous giant, $\zeta$ Hya, an odd-even pattern is still persistent, but the one sequence has a large scatter, which is very similar to  case B in the work of \cite{D-5}(2009). In an intermediate red giant branch, the convection zone is shallower, so that the coupling between the p-mode and the $g$-mode becomes strong and some mixed modes come out, in which  significant frequency shifts around the regularly spaced frequency position create scatter in an odd sequence in the echelle diagram.   In addition to this, \cite{D-5}(2009) pointed out that if the duration of observation is shorter than twice the lifetime of p-mode, the heights of the p-modes  become smaller than in the case of the longer observational duration.  In such a case, only the radial and STE modes tend to be observable.  Our results for $\epsilon$ Tau and $\eta$ Her may be explained as such a case.  However, it should be reminded that although its observational duration is sufficient, in a less-luminous giant, $\epsilon$ Oph, numerous radial modes are found, while non-radial modes with $l=1$ are a few, as presented by \cite{K-1}(2008). This also agrees very well with the theoretical prediction (smaller heights for p-modes). \\
\indent  As discussed above, our findings concerning oscillation properties in our G-type giants would be useful to understand the structural evolution of the surface convection zone from the lower part of red-giant branch to the upper part by comparing with the theoretical evolution models.  However, it should be noted that since the evolutionary tracks in the giant brach for different masses are aligned in a very narrow line, the echelle diagrams for giants at nearly the same luminosity manifest all kinds of characteristics, as pointed out by \cite{D-5}(2009).
In other words, we may determine the stellar mass from the shape of an echelle diagram. \\ 
\subsection{Damping time}
We investigate the damping time of the eigen modes for these giants, following the method proposed by \cite{S-3}(2004).  \cite{S-4}(2006) applied this method to $\xi$ Hya to estimate the damping time of the observed modes using the cumulative distribution of the power spectrum.  Since our data suffer from one-day aliasing, we followed their statistical method instead of directly measuring the width of the Lorentzian profile of the power spectrum.  \\
\indent We conducted 100 simulations of the cumulative distribution of power spectra for several damping times, ranging from 1 day to 7 days.   One example of the cumulative power distribution for $\zeta$ Hya is shown in figure 15, in which a damping time of 4 days is assumed.  One thin dotted line is the cumulative distribution of the power spectrum for one simulation, which gives the fraction (ordinate) of the cumulative power above a certain level of power (abscissa). The average distribution of these 100 cumulative distributions is shown by a thick dotted line, in which the simulated power spectrum is normalized by the observed power.  \\
\indent   We tried to find the average distribution best-fitted to  the observational cumulative power distribution, indicated by a thick solid line.  Here, it is noted that the cumulative distributions around the average line show a significantly large scatter, and therefore the accuracy of the damping time estimation is only limited to one digit or so.     The damping times ($\tau$) for 4 giants determined from such a procedure are summarized in the table 3.  \\
\indent Here, it should be reminded that  recent reports on the damping time of oscillations in red giants by \cite{D-1}(2009) and \cite{B-5}(2010) mentioned rather long damping times on the order of several tens of days from photometric observations from space with a long observation time; say, more than 100 days.  Several methods for  estimating the damping time in solar-like oscillations have been proposed by different groups, which causes, in fact, the diversity in  estimating the damping time.  In the future we should overcome such a difficulty to get a consensus on the estimated damping time.   \\
\indent Even so, it is worth while to consider the behavior of the Q-value (ratio of damping time to period) from dwarfs to giants, including our results, as previously discussed by \cite{S-4}(2006), which is shown in figure 16.   A rough trend of decrease in Q-value with increasing period is apparently seen.  If we take a close look at this figure, it is noted that the G-giant group is located in an extention of a line connecting the dwarf and subgiant groups (exceptional case of $\nu$ Ind will be discussed later). This indicates that oscillations in G-type giants are solar-like oscillations, and that their damping mechanism obeys the same rules as the Sun. If oscillations in K-type and M-type giants suffer from the same damping as the Sun, their Q-values become less than 1.  Therefore, the generated disturbances by turbulent convection cannot grow up, even to one cycle of oscillation, and thus the variations  are not recognized to be  oscillations in K-type and M-type giants. However, as shown in  figure 16, the Q-values for K-type and M-type giants are larger than 1, and also their distribution becomes flat (dotted line) against the period\footnote{In this figure, we do not plot the new results mentioned by \cite{D-1}(2009) and \cite{B-5}(2010), since the detailed data have not been released yet. However, as their points might be located above a flat dotted line, the trend we mentioned here may not change much.} . \\
\indent Here, we would suggest   two alternative interpretations on the above trend in period-Q-value diagram.  One is that the damping mechanism completely change at G-type giants, while  excitation by convection does not change, and the Q-value becomes flat, as indicated by a dotted line in figure 16.  Another one is that the excitation mechanism switches from convective excitation to excitation by the $\kappa$-mechanism, like stars in Cepheid strip, as suggested by \cite{D-2}(2001). In fact, \cite{B-3}(2005) suggested some contribution of the $\kappa$-mechanism to the excitation of oscillations in a M-type giant, $\rm{L}_2$ Pup.  \\
\indent Finally, we note that $\nu$ Ind is substantially above the  solid line in figure 16.  This star  has a  lower metallicity, [Fe/H]=  -1.4 (\cite{G-1}2000) than the rest.  This suggests  that metallicity  has an effect on the mode damping.     \\

 \section{Conclusion}
The power spectra of the time series for 4 G-type giants (11Com, $\zeta$ Hya, $\epsilon$ Tau, and $\eta$ Her) have a Gaussian shape, and their overall frequency distribution is unchanged when observed in different seasons, as shown in 11 Com and $\zeta$ Hya.   Their characteristic frequencies $\nu_{max} $ are well defined as a peak of the power distribution, which are in a good agreement with those predicted by the scaling law.   \\
\indent We modified a new method proposed by \cite{C-2}(2007) to umbiguously determine any large frequency separation by using the highest peak distribution of the collapsed power spectrum.  The resulting large separations for these giants are in a good agreement with the predicted values by the scaling law for the large separation.  \\
\indent The echelle diagrams of the oscillation frequency for 4 G-type giants constructed from  the large separations clarify the properties of  the oscillation modes in G-type giants; i.e.,  non-radial modes are detected in addition to radial modes, although in the lower luminous ($L < 100 L_{\odot}$) G-type giants, like $\epsilon$ Tau and $\eta$ Her, one sequence has numerous modes, and another has only one mode.  This fact is just what \cite{D-5}(2009) predicted theoretically on the mode properties in giants.  A detailed comparison of the observational echelle diagrams with the theoretical models is urgently needed in order to understand the structural evolution of the surface convection zone in G-type giants.  \\
\indent The damping times for our sample are in the range of 1-7 days.  The relation of Q-value to period, ranging from dwarfs to giants, strongly suggests a bi-modal relation, which would be very useful to understand the physical mechanism of mode excitation and damping.   It  also suggests that the metallicity  might have an effect on the lifetime of modes. \\
\vspace{1cm}\\
\indent We thank all of the staff members at Okayama Astrophysical Observatory for their support during the observations. We are also grateful to Dr. Stello for passing  lifetime information on stars, SV Lyn and R Dor.  We also thank an unanimous referee for helpful comments and suggestions leading to much improvement.  H.A. is supported by Grant-in-Aid for Scientific Research (B) (No. 17340056) from Japan Society for the Promotion of Science.  

\newpage


\newpage
\begin{longtable}{lcc}
  \caption{Observation journal}
  \hline  \hline
 Star name & Obs. period & Spectra  \\  \hline

 \hline
   11 Com & 2008.04.09 - 04.17 & 80 \\
    & 2009.03.14 - 03.20 & 215 \\
  $\zeta$ Hya & 2006.01.20 - 01.27 & 521  \\
    & 2008.12.06 - 12.14 & 243 \\ 
  $\epsilon$ Tau & 2008.12.06 - 12.14 & 323 \\
   $\eta$ Her & 2007.05.23 - 05.31 & 222 \\
  \hline
 \endhead
\end{longtable}
\begin{longtable}{lccccc} 
  \caption{Stellar parameters } 
  \hline        \hline
  Star name& $Te$ & $\rm{log} \hspace{1pt} \it{g}$  & $ L/L_{\odot}  $ & $R/R_{\odot}^{*}$ & Ref. \\ 
 \hline
\endfirsthead
\endlastfoot       
 \hline
 11 Com & 4742 & 2.31 &  172 &  19.5   & 1  \\
 $\zeta$ Hya & 4844 & 2.48** &  132 &  17.9  &  3  \\
 $\epsilon$ Tau & 4901 & 2.64 &  97 &  13.7  &  2  \\
 $\eta$ Her & 5045 & 2.79 &  50 &  9.2   &  3 \\
  \hline
  \multicolumn{6}{l}{* Radii are calculated from Hipparcos parallaxes and interferometrically}\\
   \multicolumn{6}{l}{ determined apparent diameters (see the text) except for 11Com.} \\ 
  \multicolumn{6}{l}{** modified from \cite{T-1}(2008). See the text for the details.}  \\
  \multicolumn{6}{l}{1: \cite{S-5}(2008), 2: \cite{S-6}(2007), 3: \cite{T-1}(2008)}
  \end{longtable}
\begin{longtable}{lccccc}
  \caption{Characteristic Frequencies}
  \hline        \hline
 Star name & $\nu_{max}$($\mu$Hz) &  $\nu_{max,sc}$ & $\triangle \nu$ ( $\mu$Hz) & $\triangle \nu_{sc}$  & $\tau$(days)  \\  
 \hline
  11 Com & 29.6 & 25.0 & - &  2.63 & -\\
    & 26.7 & - & 2.88 & - & 7 \\
 $\zeta$ Hya & 34.1 & 36.5 & 3.26 & 3.34 & 4  \\
    & 34.6 & - & - & - & - \\ 
  $\epsilon$ Tau & 62.0 & 52.5 & 4.18 & 4.59 & 1 \\
  $\eta$ Her & 83.0 & 73.1 & 5.76 & 6.65 & 5 \\
  \hline
 \multicolumn{6}{l}{$\nu_{max,sc}$ and $\triangle \nu_{sc}$ are calculated from respective scaling laws.}    \\
  \endhead
\end{longtable}  
\begin{longtable}{cccccc}
  \caption{Extracted frequencies of 11 Com}
  \hline  \hline
  \multicolumn{3}{c}{2008} & \multicolumn{3}{c}{2009}  \\       
  Freq.($\mu$Hz) & Amp.($\rm{ms}^{-1}$)  & S/N & Freq.($\mu$Hz) & Amp.($\rm{ms}^{-1}$)  & S/N \\  \hline
\endhead
\endfoot 
 \hline 
\endlastfoot        
 \hline
 - & - & - & 9.40 & 2.2 & 4.4  \\
  13.6 & 2.4 & 3.4 & - & - & -  \\
  - & - & - & 15.4 & 5.7 & 11.7  \\
  - & - & - & 16.7 & 3.7 & 7.7  \\
  17.8 & 3.1 & 4.4 & - & - & -  \\ 
  27.0 & 7.5 & 10.7 & - & - & - \\
  - & - & - & 29.8 & 1.5 & 3.0  \\
  - & - & - & 34.0 & 3.5 & 7.1  \\
  - & - & - & 35.5 & 5.3 & 10.9  \\
  41.1 & 2.9 & 4.1 & - & - & - \\
  - & - & - & 68.5 & 1.6 & 3.2  \\
\end{longtable}
\begin{longtable}{cccccc}
  \caption{Extracted frequencies of $\zeta$ Hya}
  \hline  \hline
  \multicolumn{3}{c}{2006} & \multicolumn{3}{c}{2008}  \\       
  Freq.($\mu$Hz) & Amp.($\rm{ms}^{-1}$)  & S/N & Freq.($\mu$Hz) & Amp.($\rm{ms}^{-1}$)  & S/N \\  \hline
\endhead
\endfoot 
 \hline 
\endlastfoot        
 \hline
  3.08 & 2.4 & 8.7 & - & - & -  \\
  10.2 & 2.0 & 7.2 & - & - & -  \\ 
  - & - & - & 17.7 & 2.7 & 3.1  \\
  25.5 & 1.1 & 4.1 & - & - & - \\
  29.9 & 4.4 & 16.1 & - & - & -  \\
  32.1 & 4.5 & 16.4 & 32.5 & 7.5 & 8.5  \\
  33.8 & 2.9 & 10.6 & - & - & -  \\
  38.6 & 3.3 & 12.1 & - & - & -  \\
  58.9 & 2.0 & 7.4 & 59.7 & 2.7 & 3.1 \\
  75.5 & 1.1 & 3.9 & - & - & -  \\
  80.9 & 0.9 & 3.2 & - & - & -  \\
\end{longtable}
\begin{longtable}{ccc}
  \caption{Extracted frequencies of $\epsilon$ Tau}
  \hline  \hline
  \multicolumn{3}{c}{2008}   \\       
  Freq.($\mu$Hz) & Amp.($\rm{ms}^{-1}$) & S/N  \\  \hline
\endhead
\endfoot 
 \hline 
\endlastfoot        
 \hline
  10.1 & 1.4 & 3.3    \\
  33.1 & 2.5 & 6.1  \\ 
  51.9 & 3.4 & 8.2  \\
  64.5 & 4.0 & 9.7   \\
  72.7 & 1.3 & 3.2   \\
  90.2 & 2.0 & 4.9  \\
 \end{longtable}
\begin{longtable}{ccc}
  \caption{Extracted frequencies of $\eta$ Her}
  \hline  \hline
  \multicolumn{3}{c}{2007}   \\       
  Freq.($\mu$Hz) & Amp.($\rm{ms}^{-1}$) & S/N  \\  \hline
\endhead
\endfoot 
 \hline 
\endlastfoot        
 \hline
  75.6 & 1.8 & 3.3    \\
  83.8 & 4.2 & 7.5  \\ 
  90.0 & 2.3 & 4.1  \\
  118.7 & 1.9 & 3.4   \\
 \end{longtable}    
\newpage
\begin{figure}
\includegraphics[width=10cm,clip]{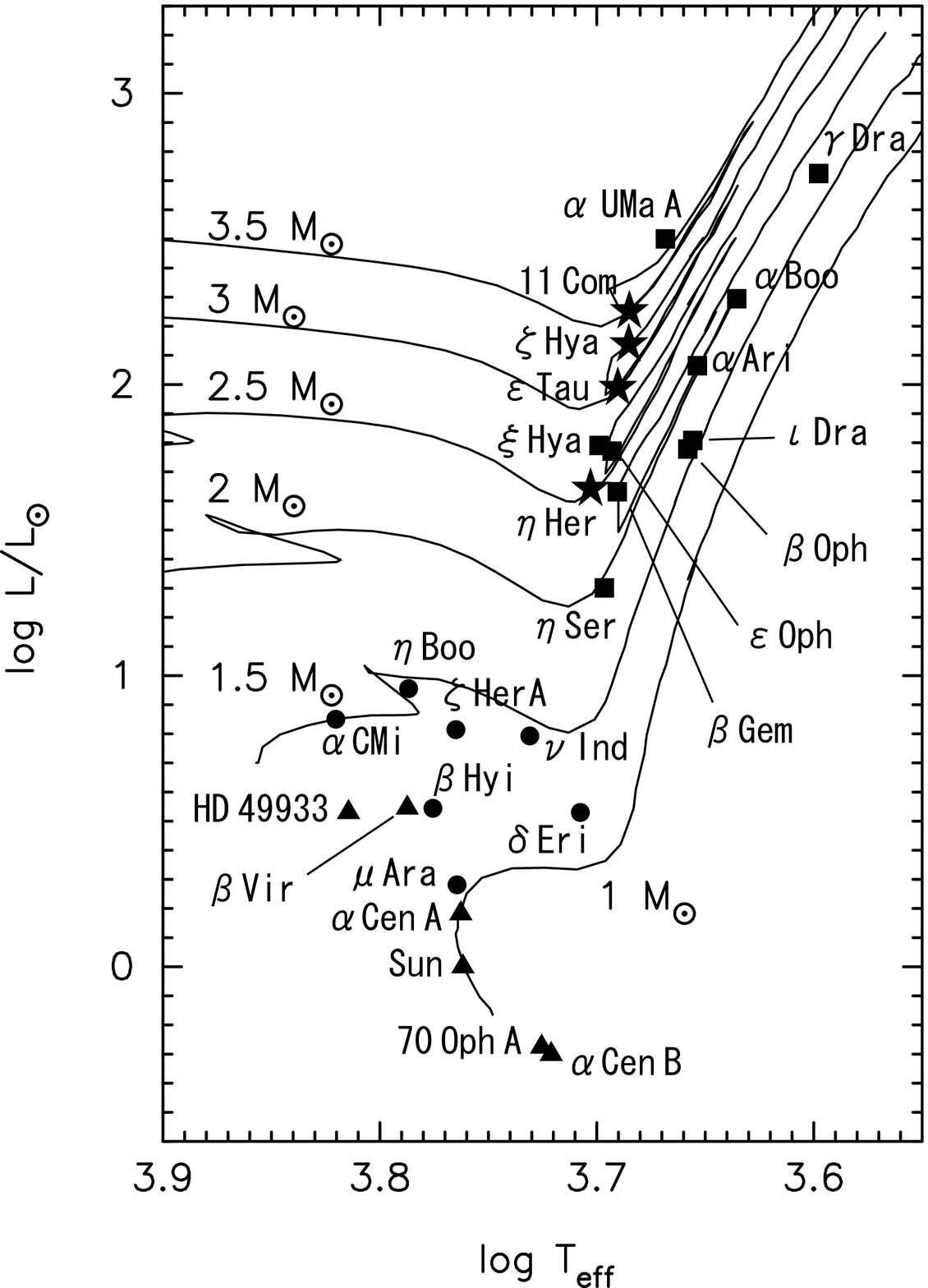} \\
 Fig. 1  H-R diagram with our G-type giants (filled star symbols).  \\
\hspace{11mm} Additional stars that show  evidence of solar-like oscillations \\
\hspace{11mm} are marked (dwarfs: filled triangles, subgiants: filled circles, giants: filled squares). \\
\hspace{11mm} Solid lines are evolution tracks for solar metallicity stars calculated \\
\hspace{11mm} by \cite{G-4}(2000).   
\end{figure}
\begin{figure}
\includegraphics[width=10cm,clip]{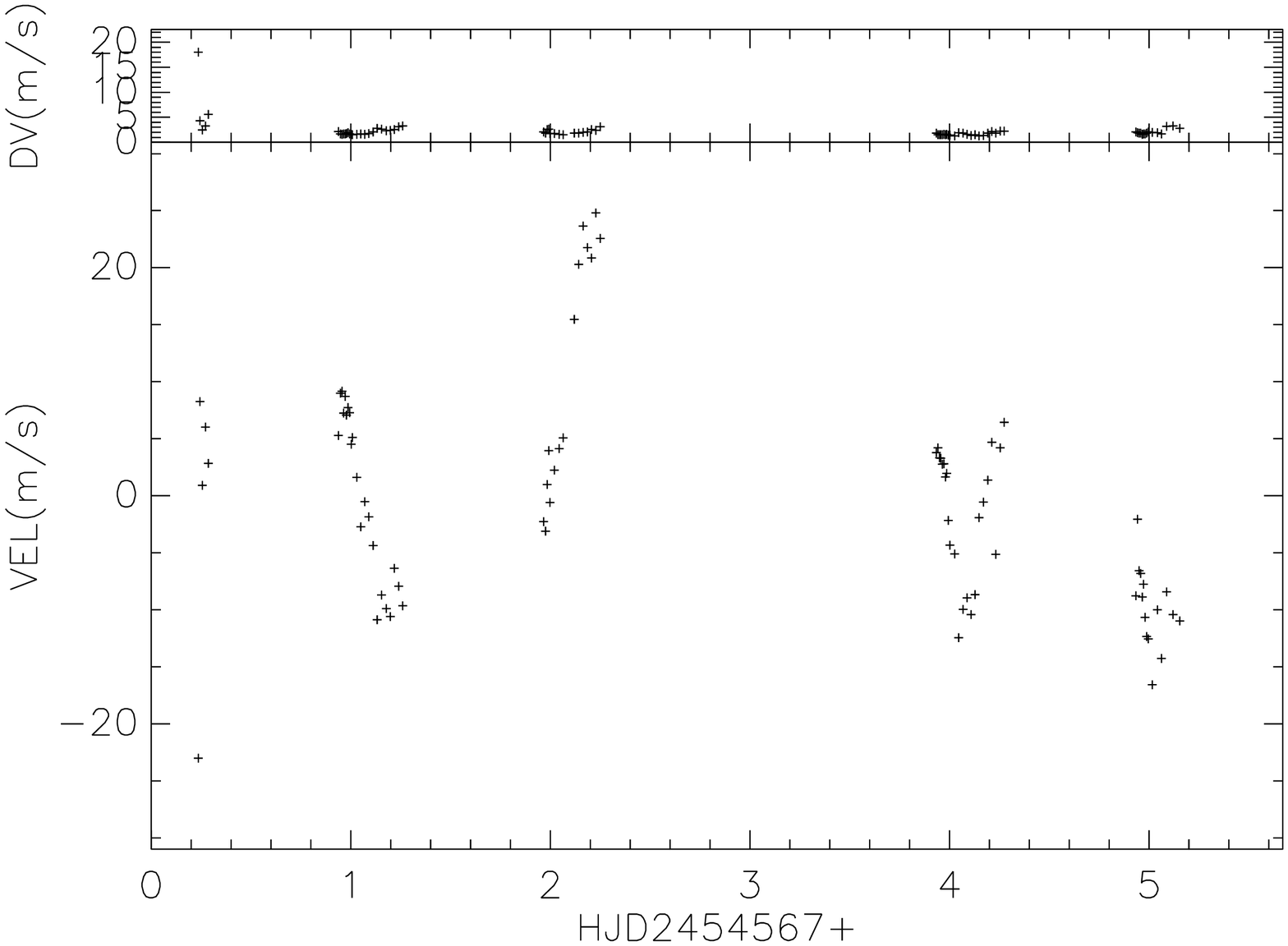} \\
 Fig. 2a  Radial-velocity  variation of 11 Com  in 2008. \\
 \hspace{13mm} Upper panel shows the internal error for each point.   
\end{figure}
\begin{figure}
\includegraphics[width=10cm,clip]{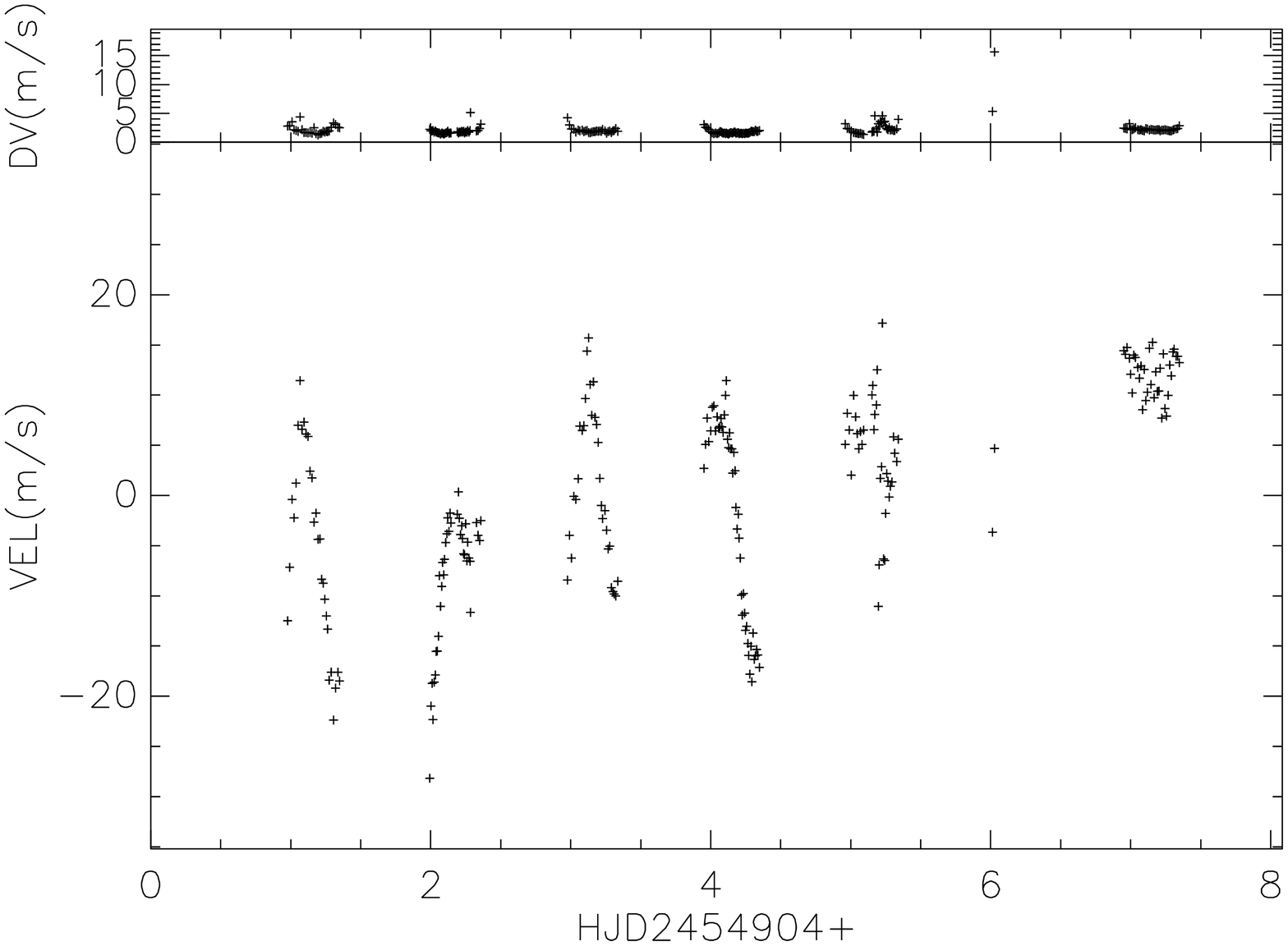} \\
 Fig. 2b  Radial-velocity  variation of 11 Com  in 2009 \\
\hspace{13mm} Time coverage is better than that in 2008. \\
\hspace{13mm} But the time variation is very similar.   
\end{figure}
\begin{figure}
\includegraphics[width=10cm,clip]{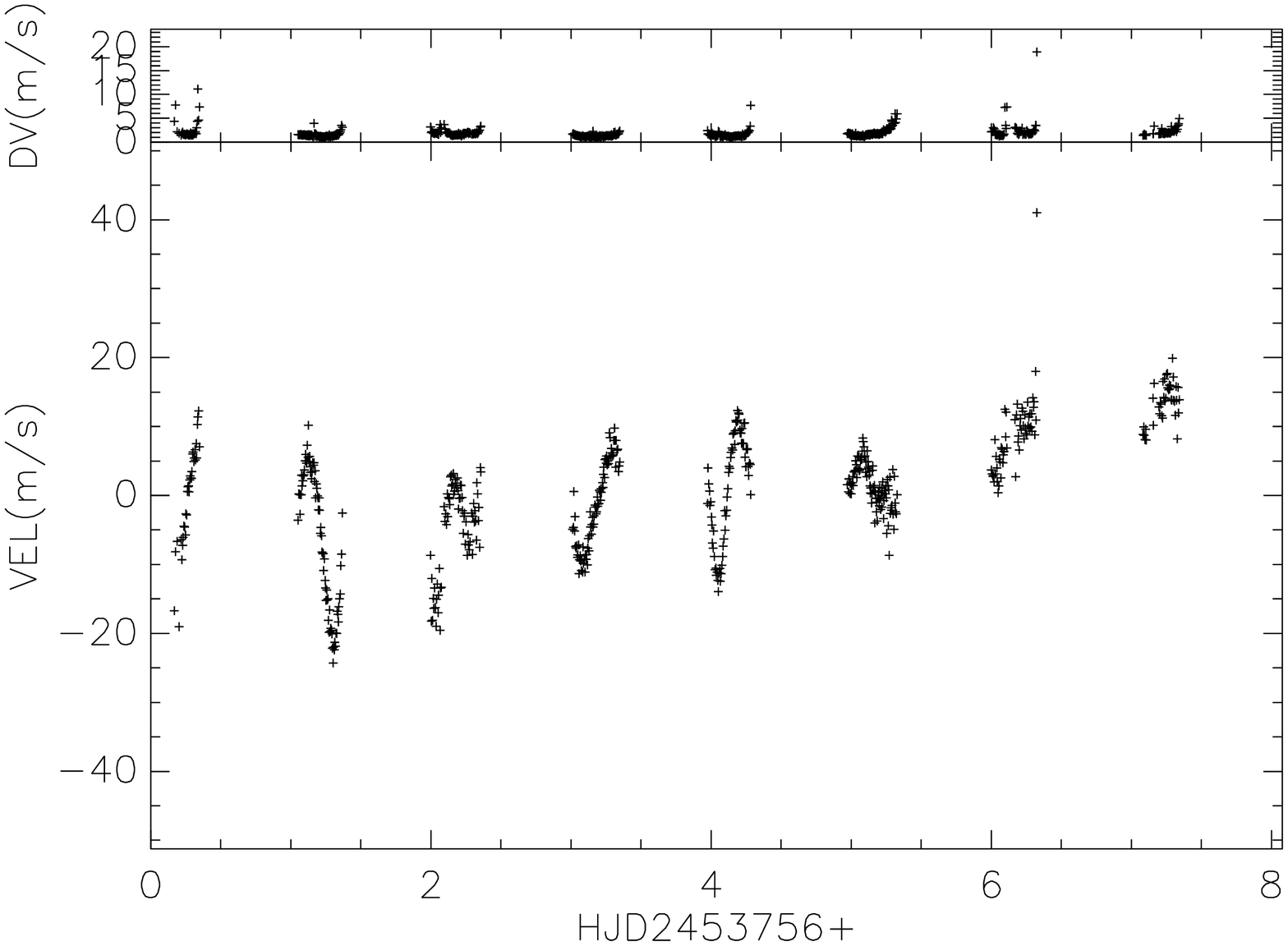} \\
 Fig. 3a  Radial-velocity  variation of $\zeta$ Hya in 2006 
\end{figure}
\begin{figure}
\includegraphics[width=10cm,clip]{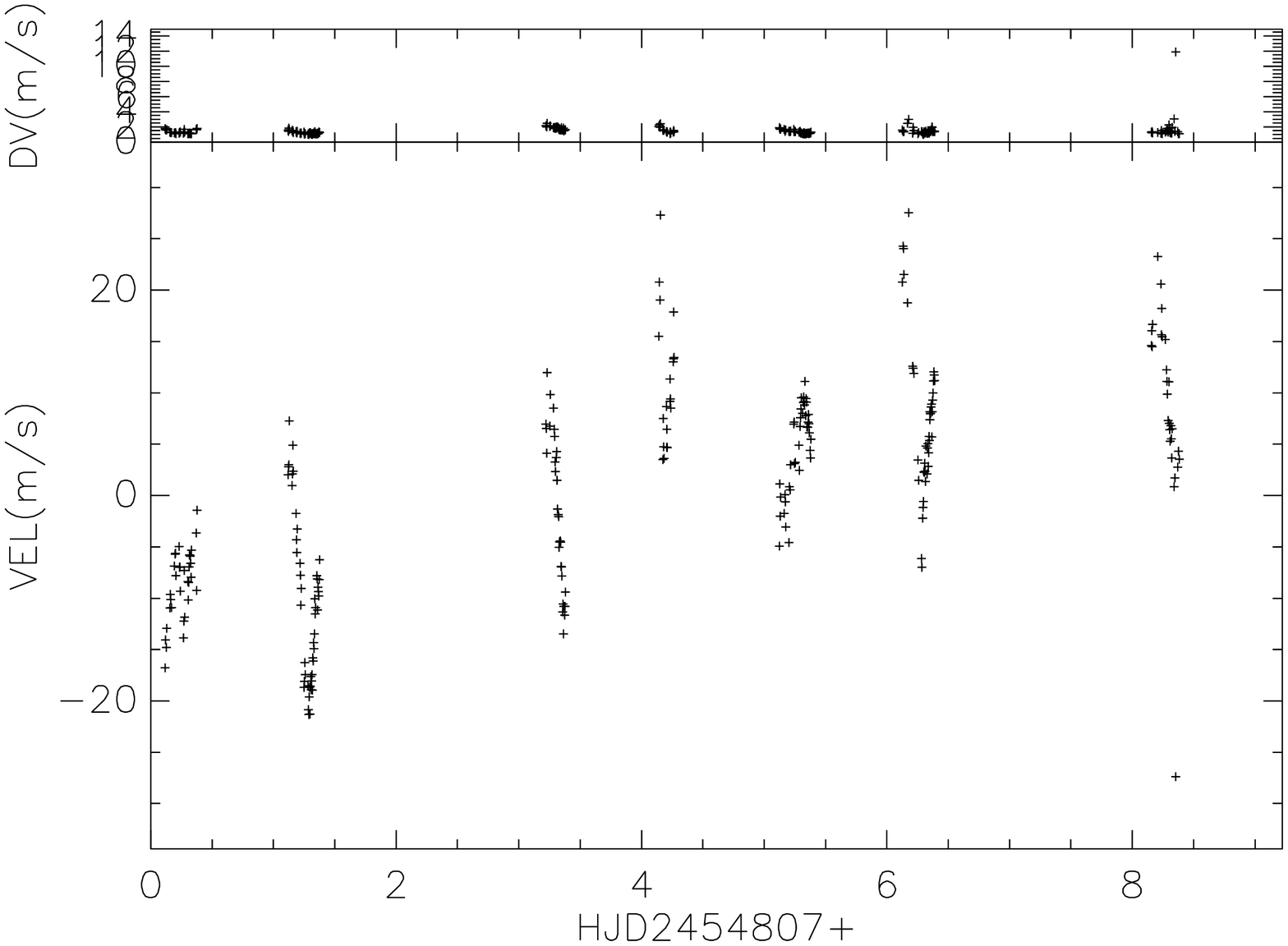} \\
Fig. 3b  Radial-velocity  variation of $\zeta$ Hya in 2008 
\end{figure}
\begin{figure}
\includegraphics[width=10cm,clip]{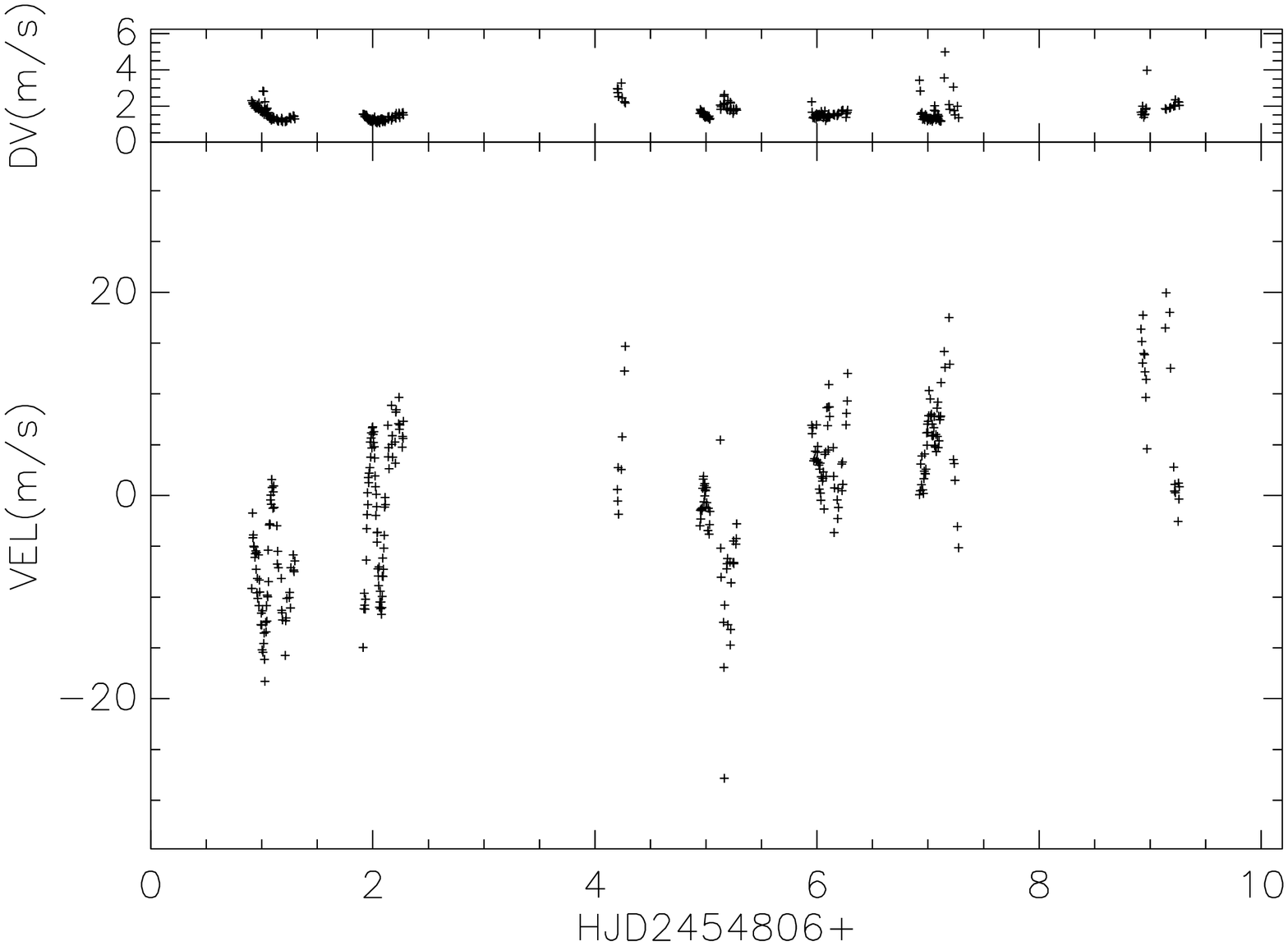} \\
 Fig. 4  Radial-velocity  variation of $\epsilon$ Tau  in 2008 
\end{figure}
\begin{figure}
\includegraphics[width=10cm,clip]{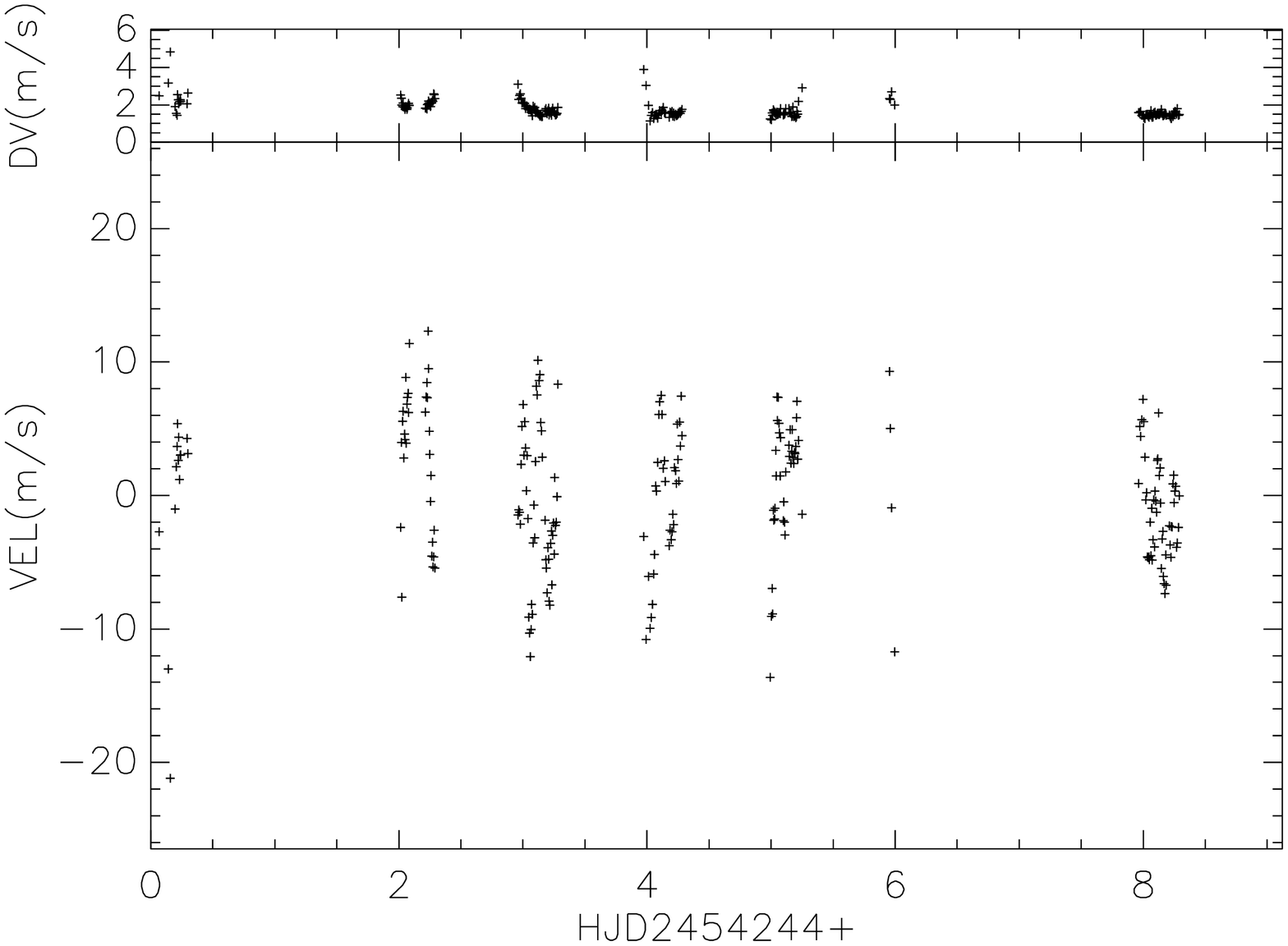} \\
 Fig. 5  Radial-velocity  variation of $\eta$ Her  in 2007 
\end{figure}
\begin{figure}
\includegraphics[width=10cm,clip]{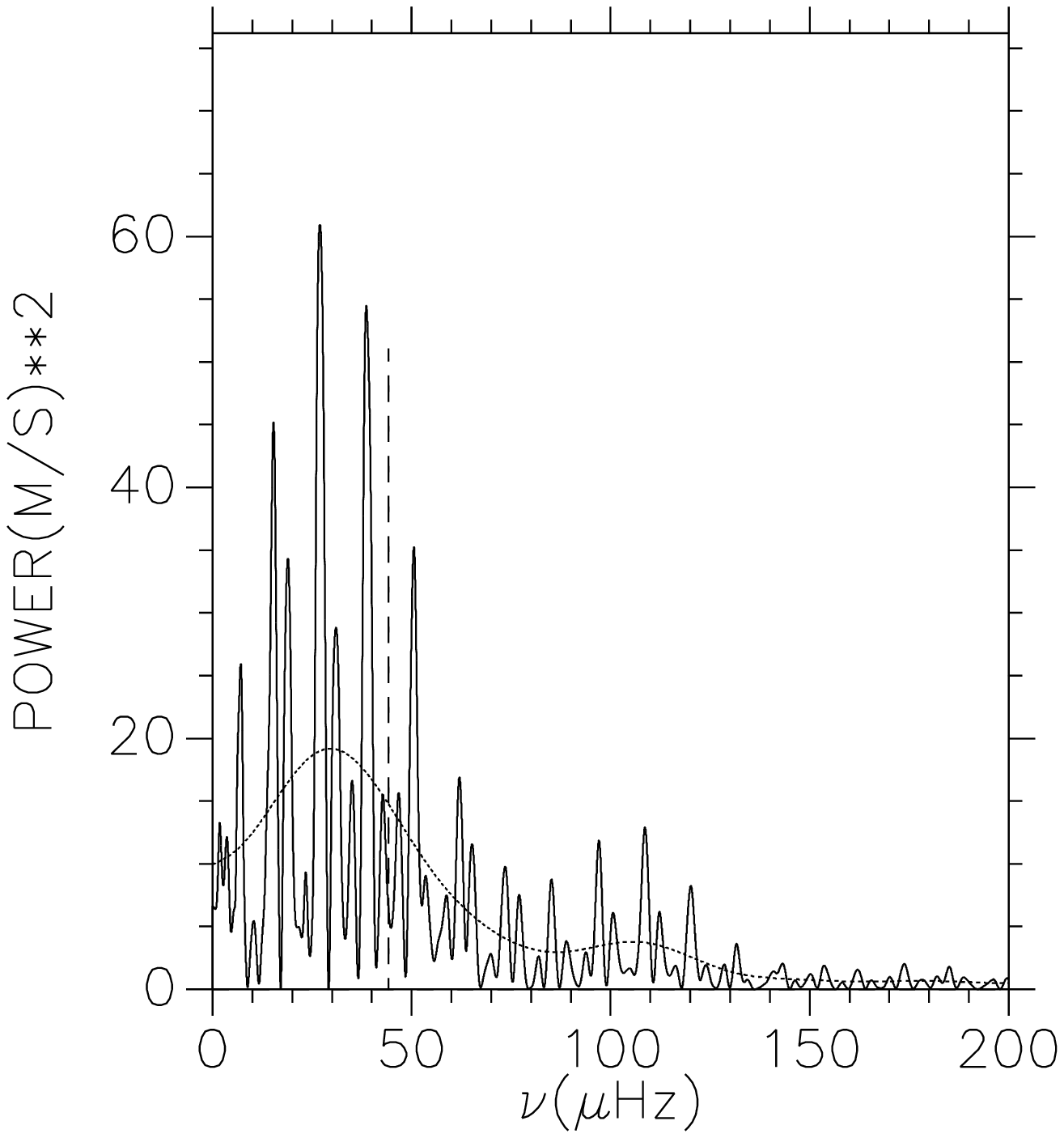} \\
 Fig. 6a  Power spectrum distribution  of 11Com  in 2008. \\
 \hspace{13mm} Vertical dashed line shows position of acoustic cut-off frequency. 
\end{figure}
\begin{figure}
\includegraphics[width=10cm,clip]{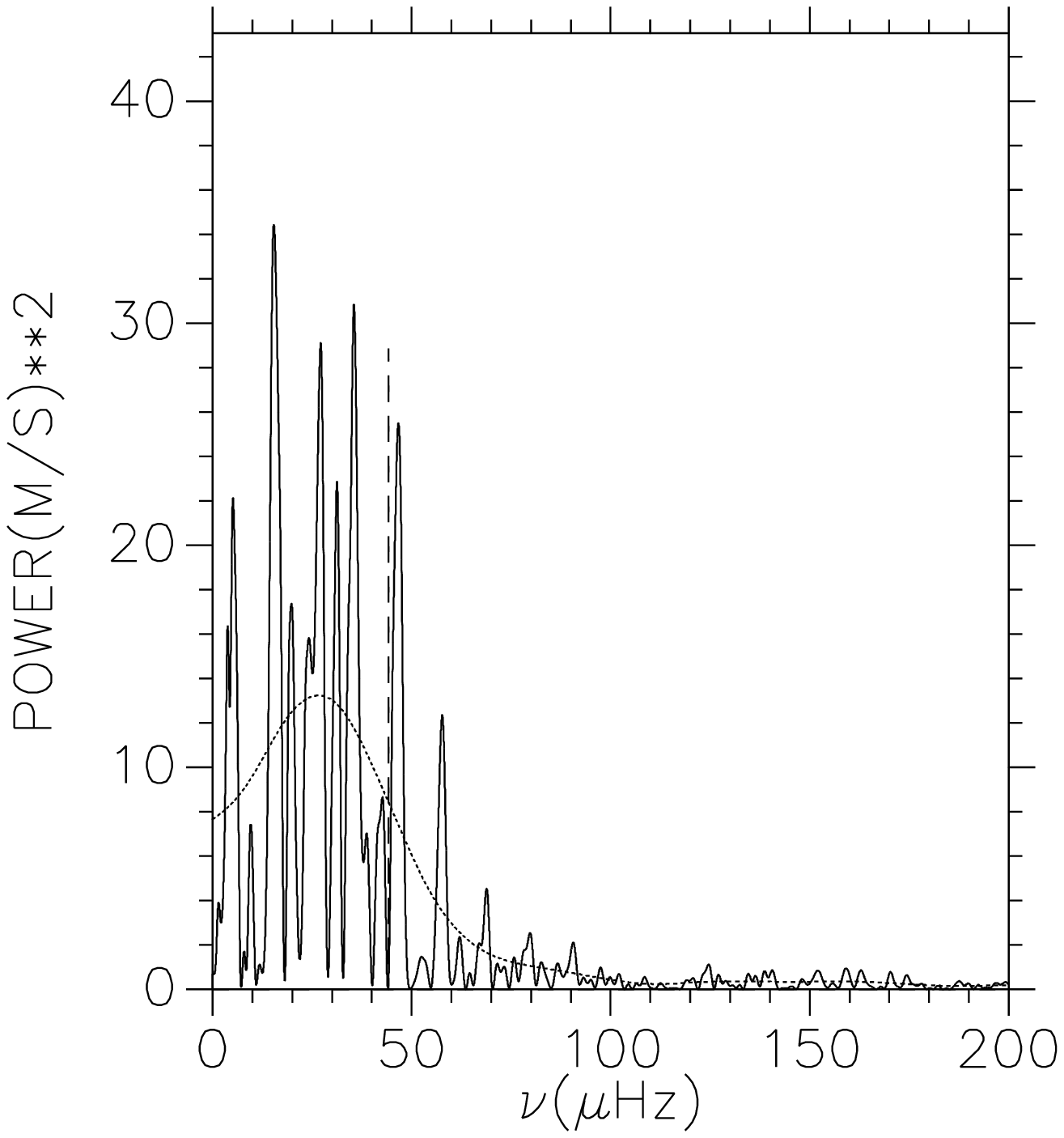} \\
 Fig. 6b  Power spectrum distribution  of 11Com  in 2009 
\end{figure}
\begin{figure}
\includegraphics[width=10cm,clip]{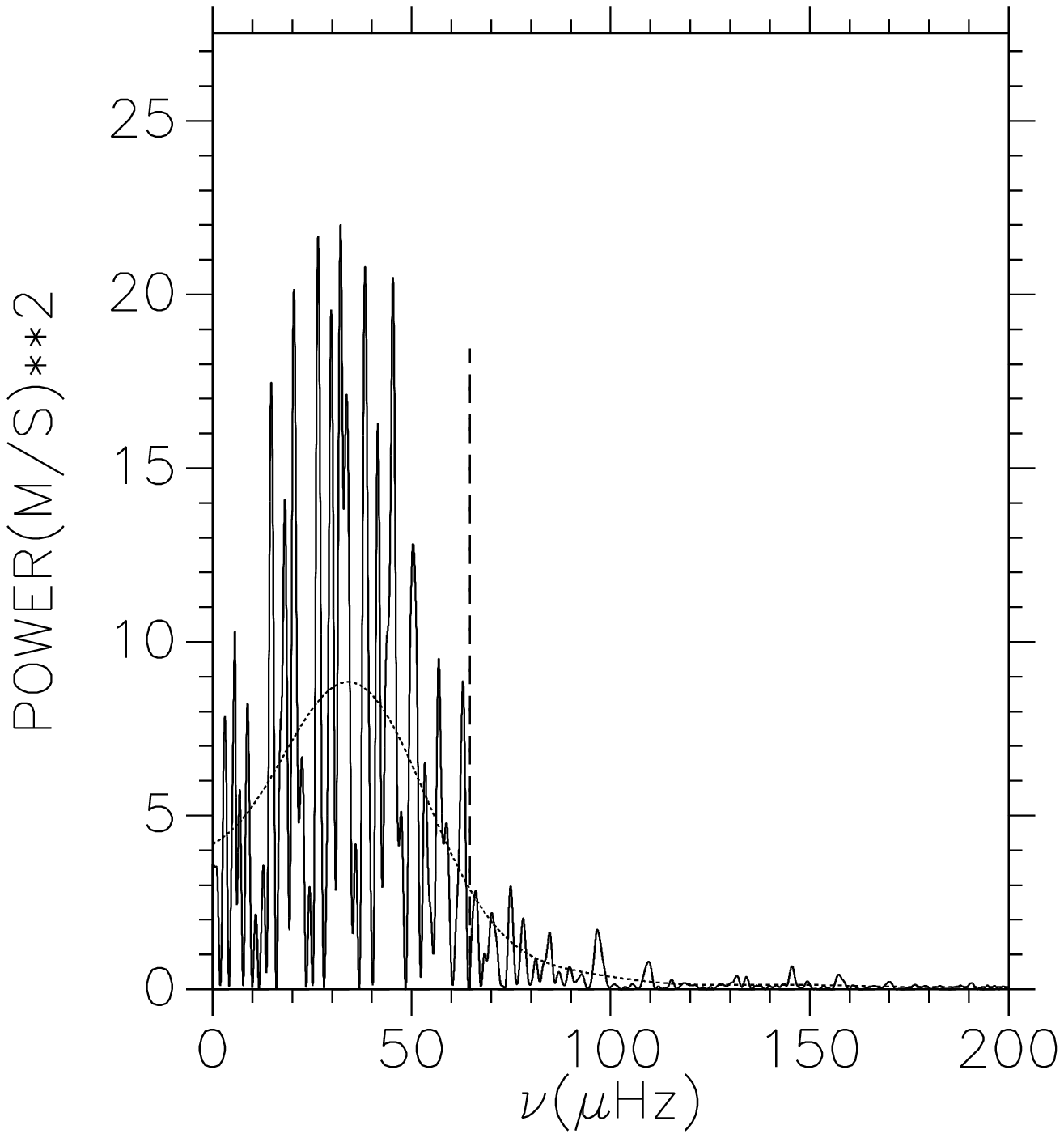} \\
 Fig. 7a  Power spectrum distribution  of $\zeta$ Hya  in 2006 
\end{figure}
\begin{figure}
\includegraphics[width=10cm,clip]{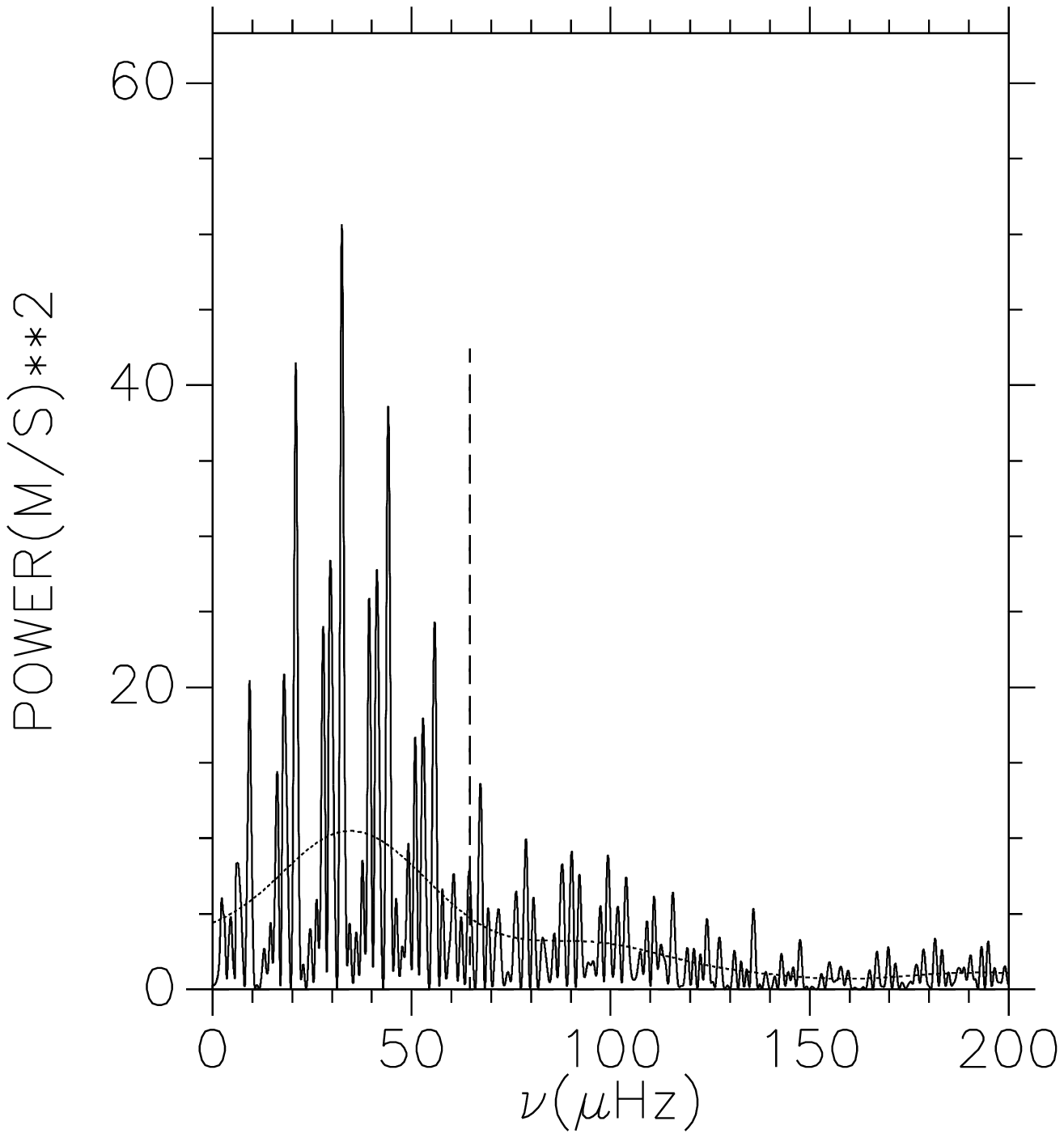} \\
 Fig. 7b  Power spectrum distribution  of $\zeta$ Hya  in 2008 
\end{figure}
\begin{figure}
\includegraphics[width=10cm,clip]{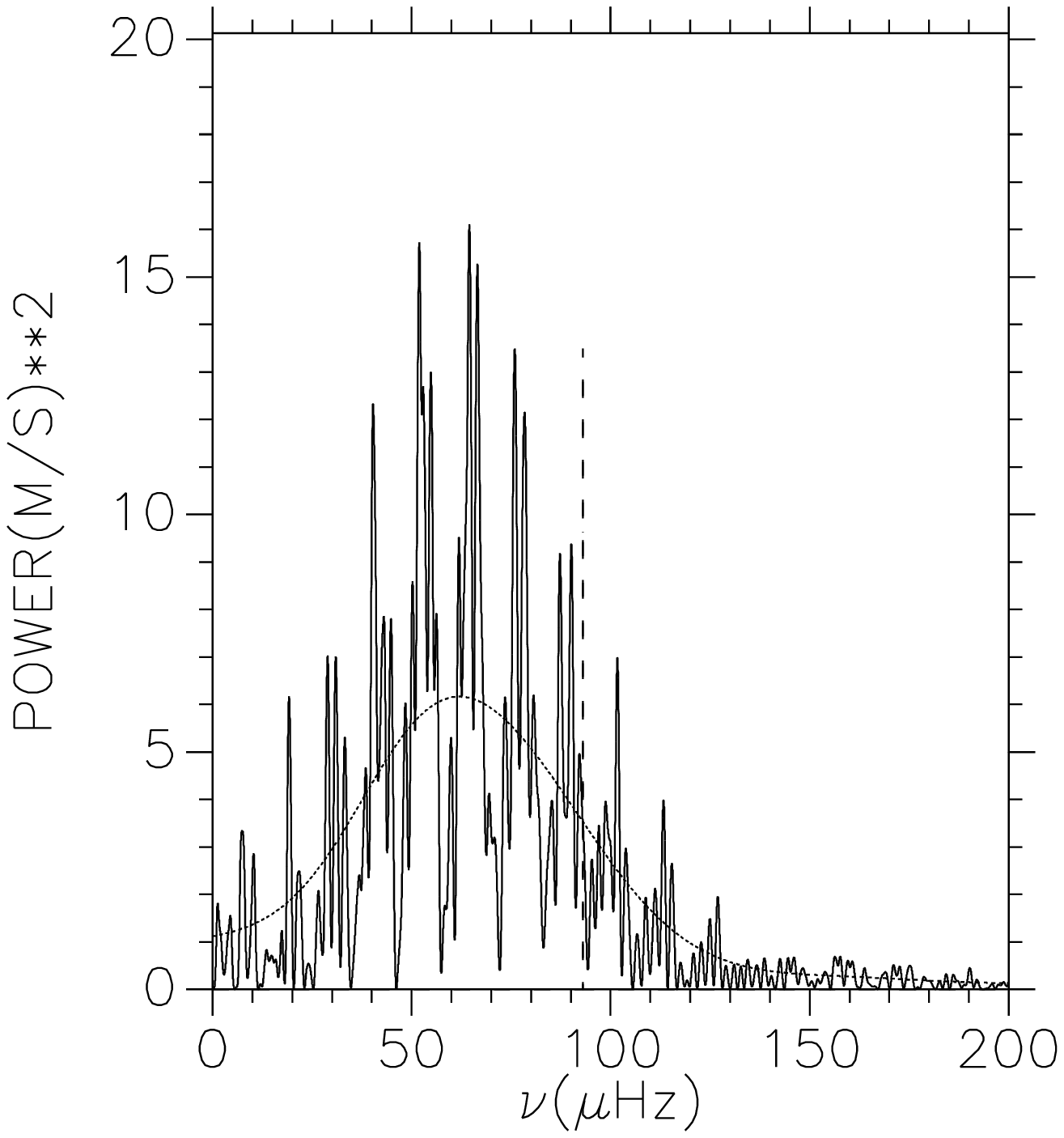} \\
 Fig. 8  Power spectrum distribution  of $\epsilon$ Tau  in 2008 
\end{figure}
\begin{figure}
\includegraphics[width=10cm,clip]{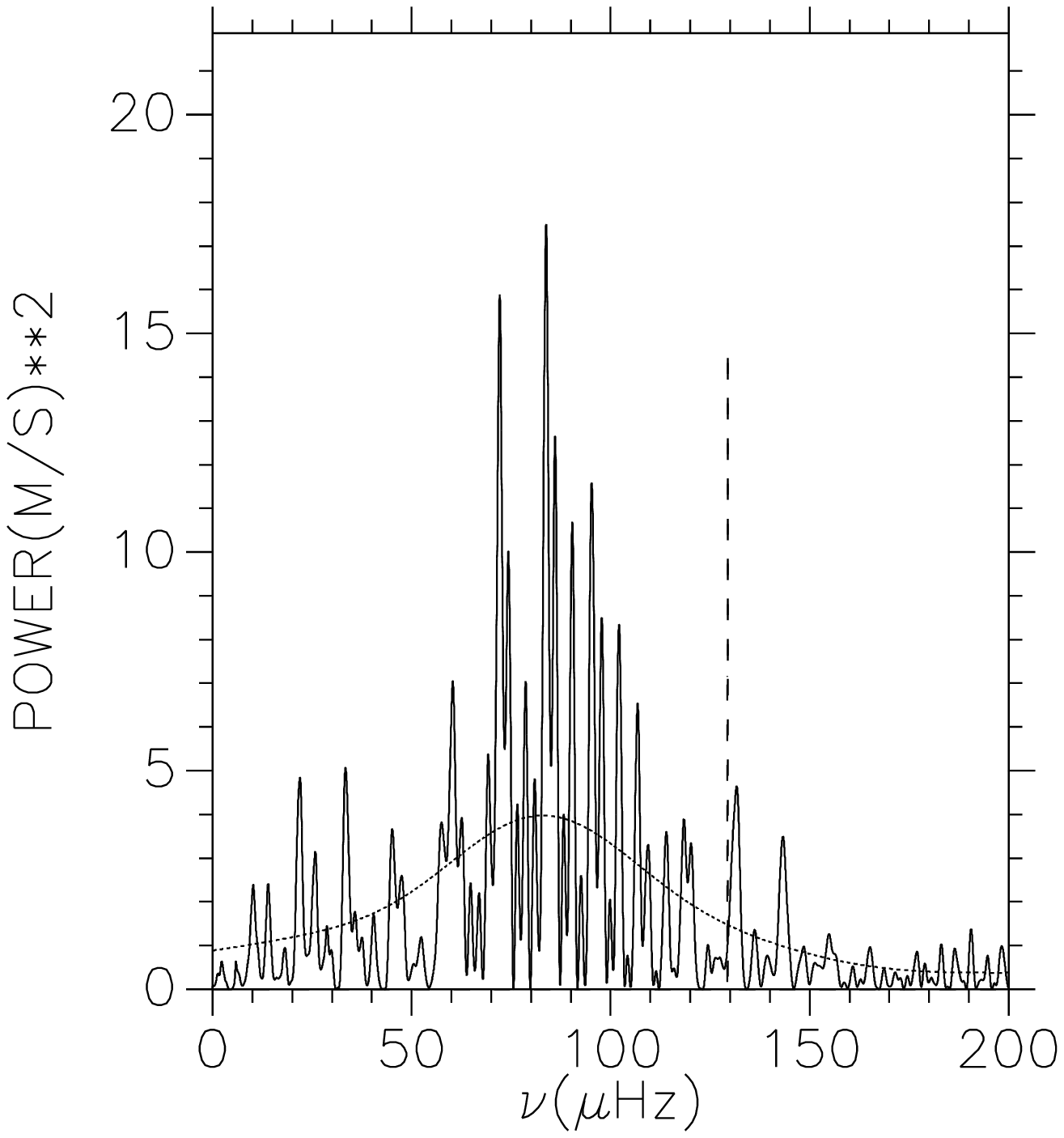} \\
 Fig. 9  Power spectrum distribution  of $\eta$ Her  in 2007 
\end{figure}
\begin{figure}
\includegraphics[width=10cm,clip]{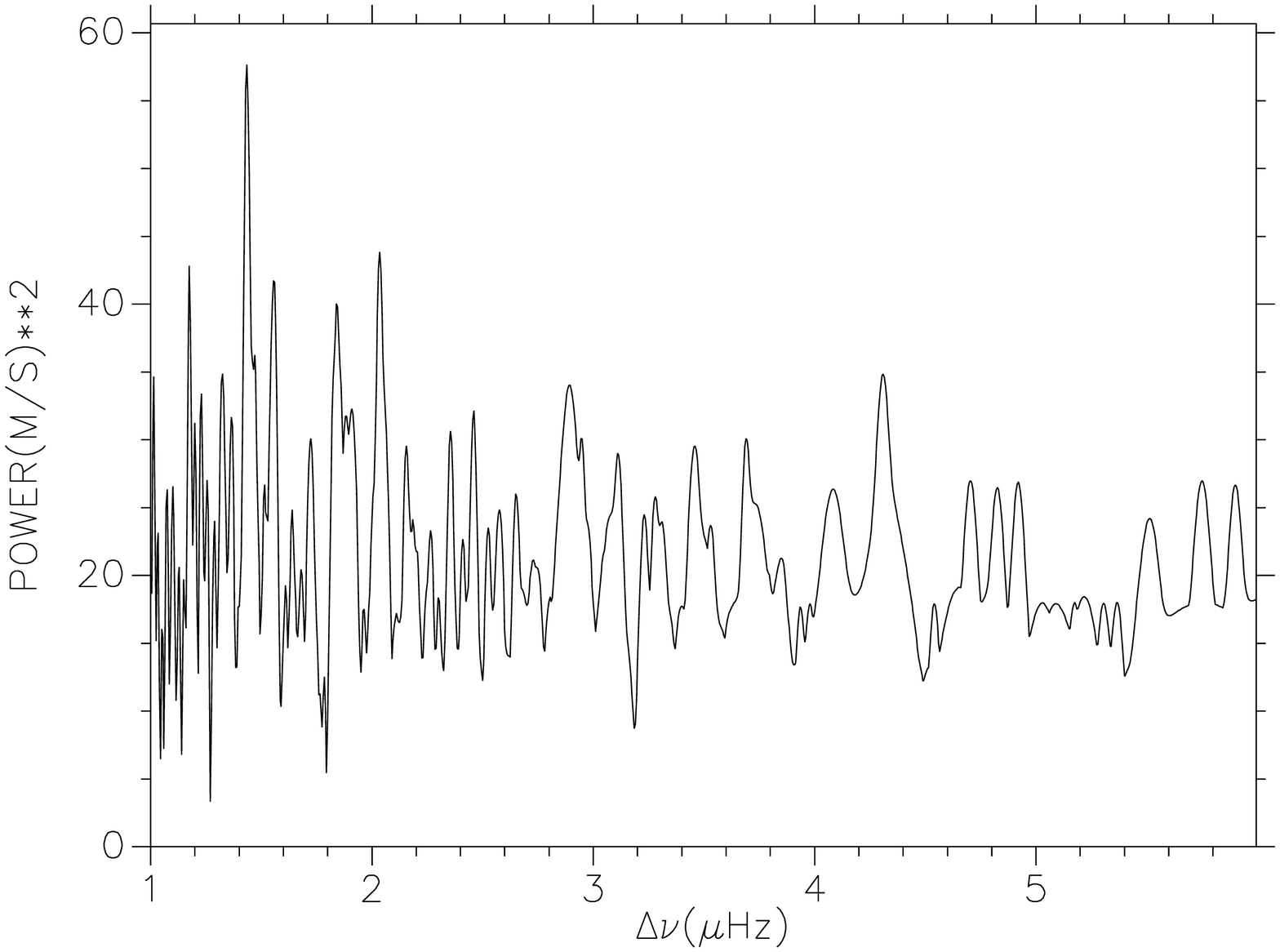} \\
 Fig. 10  Distribution of highest peak in the collapsed power spectrum  for 11 Com(2009).  \\
 \hspace{13mm} The ordinate is arbitrary unit. The highest peak is seen at 1.44 $\mu$Hz. \\
 \hspace{13mm} As a consequence, it gives 2.88 $\mu$Hz for the large separation $\triangle \nu$ of 11 Com(2009).
\end{figure}
\begin{figure}
\includegraphics[width=10cm,clip]{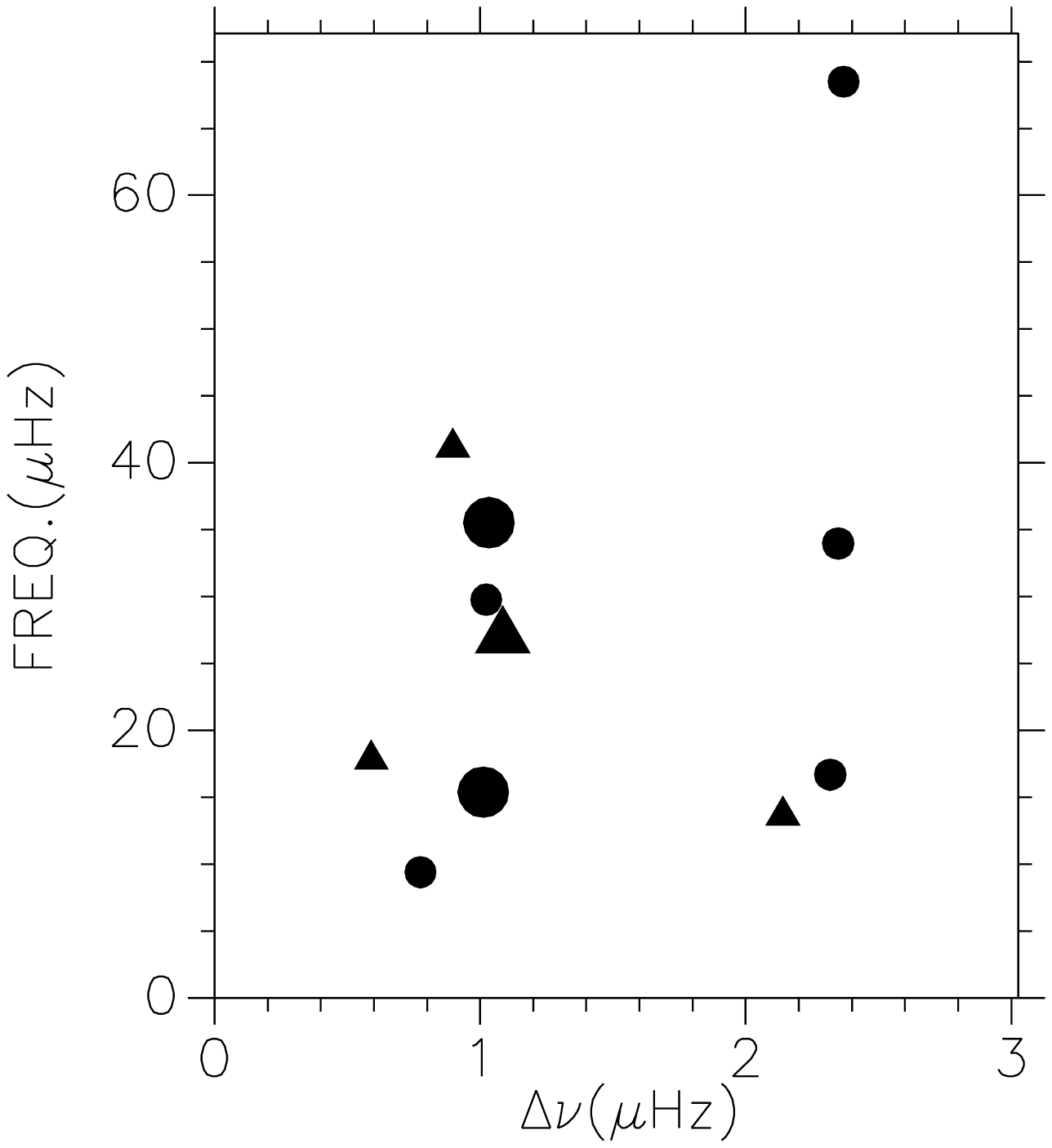} \\
 Fig. 11  Echelle diagram of frequencies for 11 Com.  \\
 \hspace{13mm} Filled circles for 2009 and filled triangles for 2008. \\
 \hspace{13mm} Small mark for a peak with amplitude ranging from 3$\sigma$ to 10$\sigma$. \\
 \hspace{13mm} Large mark for a peak with amplitude above 10$\sigma$.
\end{figure}
\begin{figure}
\includegraphics[width=10cm,clip]{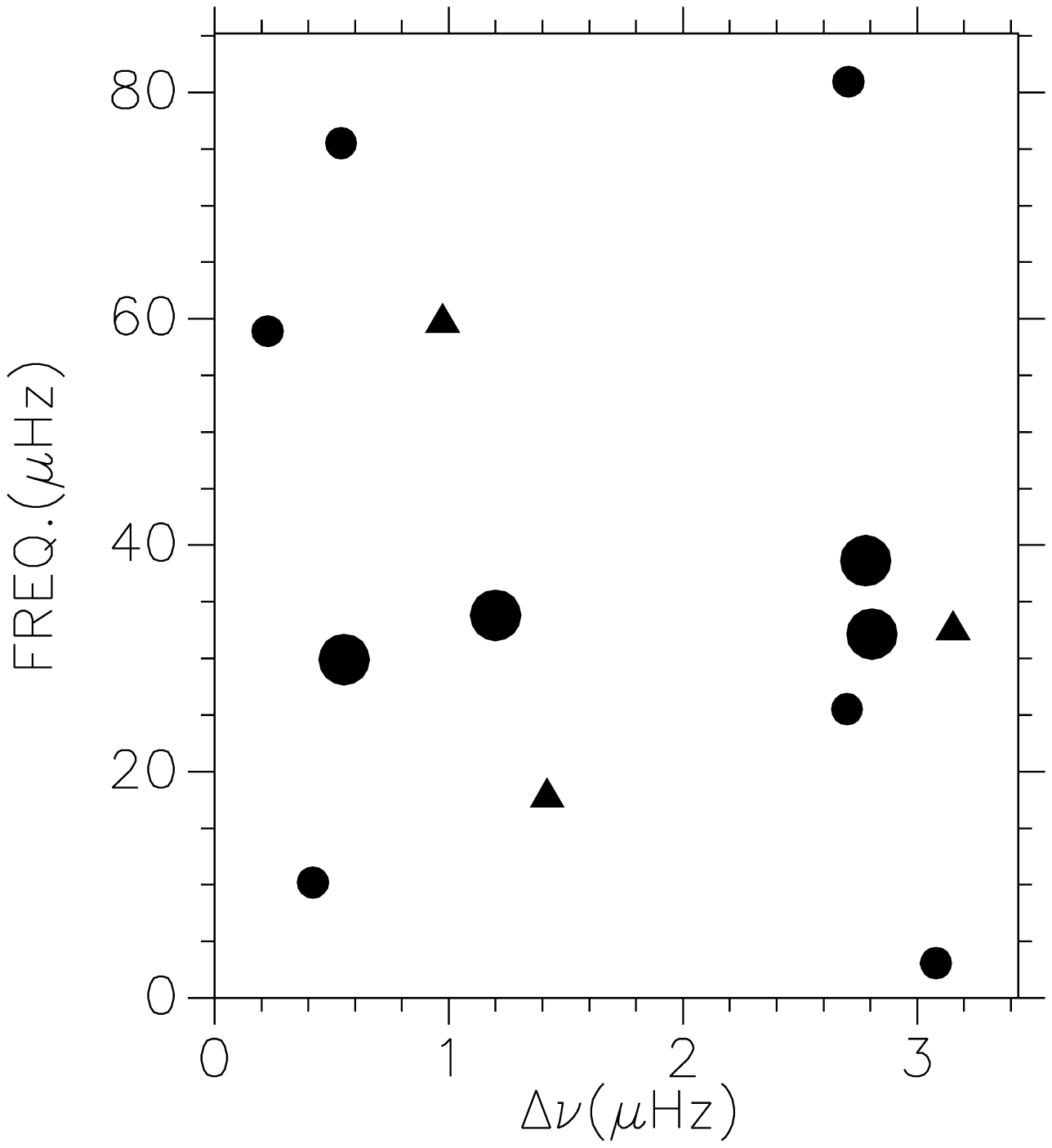} \\
 Fig. 12  Echelle diagram of frequencies for $\zeta$ Hya.  \\
 \hspace{13mm} Filled circles for 2006 and filled triangles for 2008. \\
  \hspace{13mm} Marks have the same meanings as in figure11.
\end{figure}
\begin{figure}
\includegraphics[width=10cm,clip]{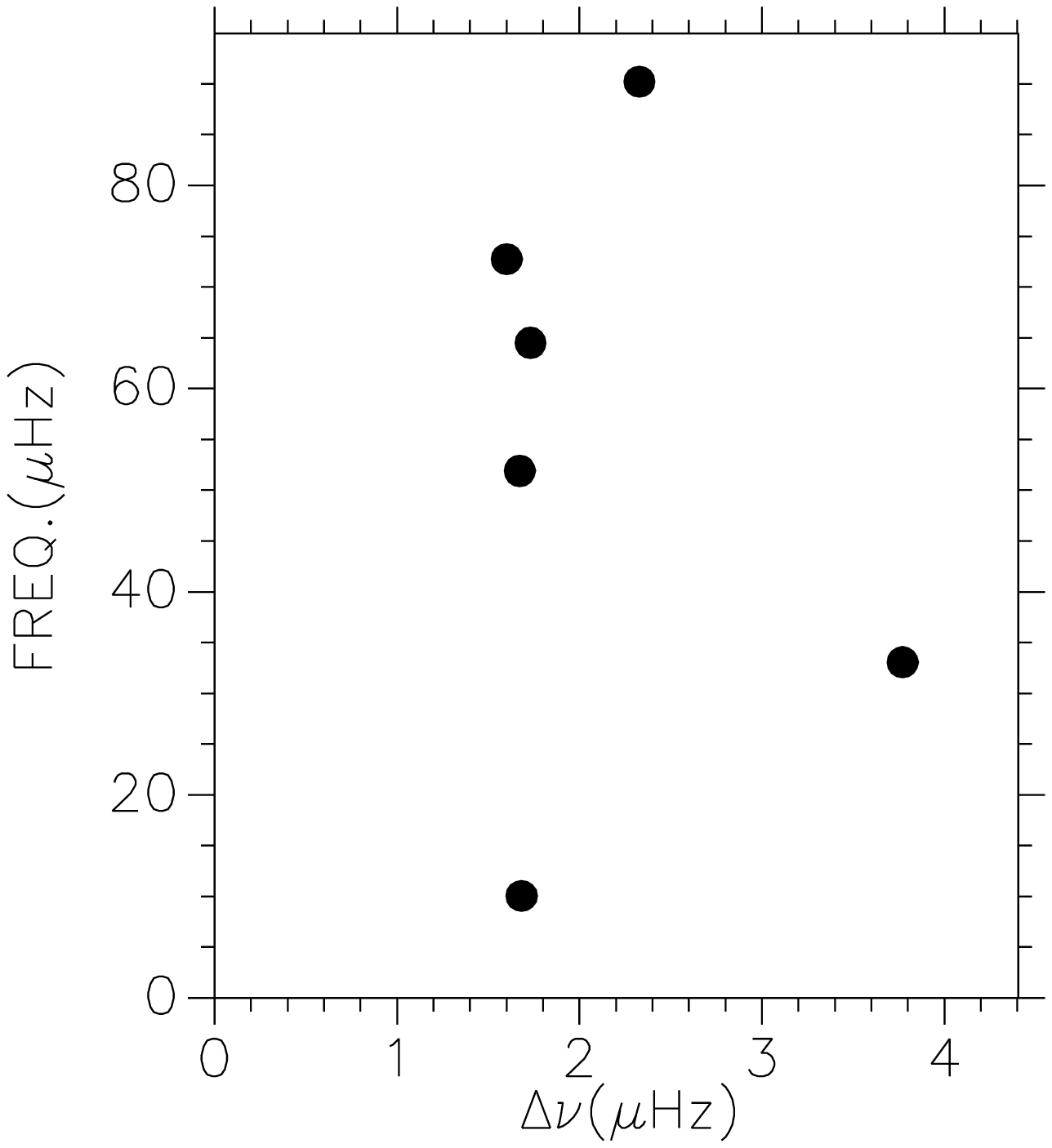} \\
 Fig. 13  Echelle diagram of frequencies for $\epsilon$ Tau.  \\
 \hspace{13mm} Marks have the same meanings as in figure11.
\end{figure}
\begin{figure}
\includegraphics[width=10cm,clip]{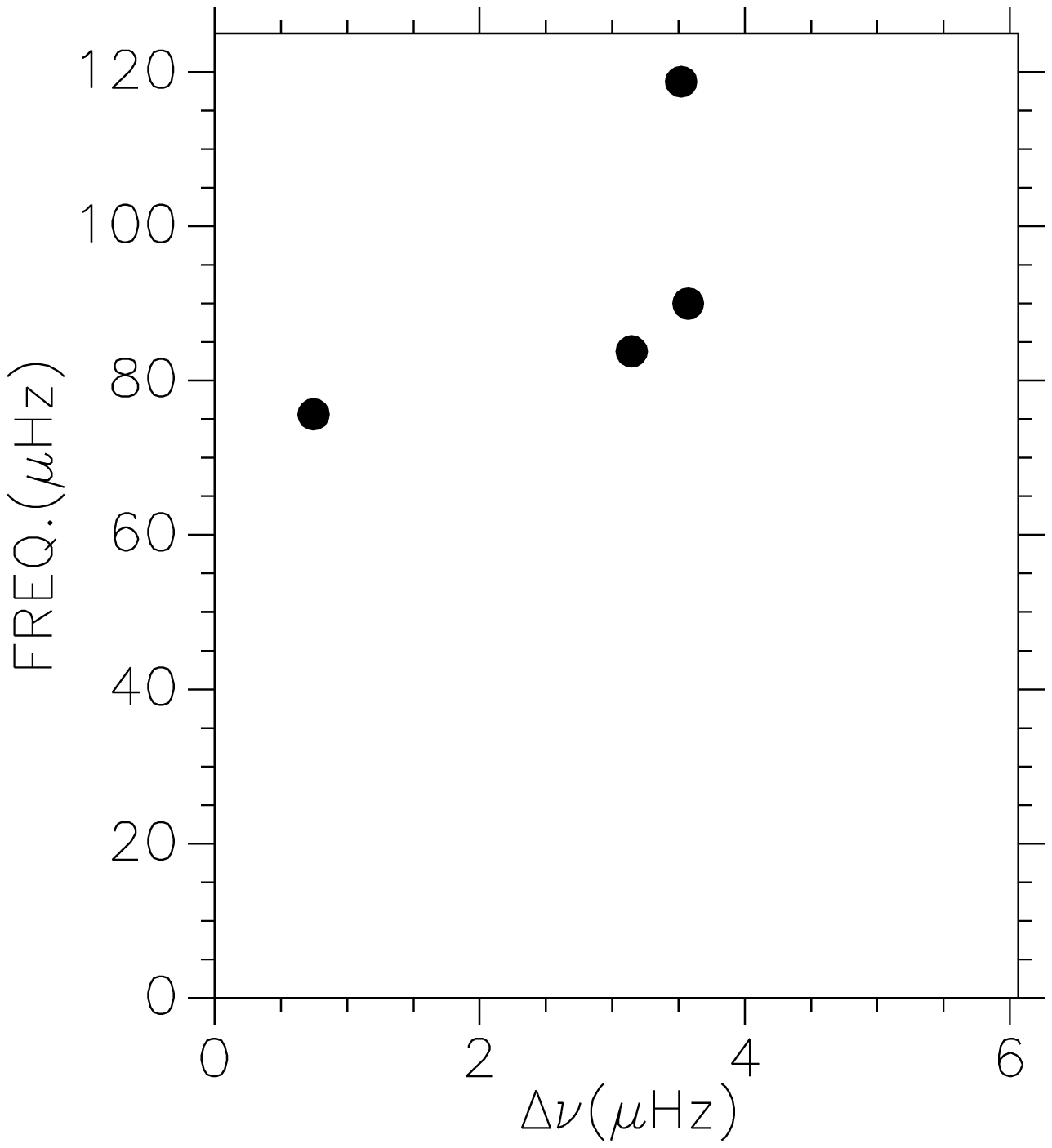} \\
 Fig. 14  Echelle diagram of frequencies for $\eta$ Her.  \\
 \hspace{13mm} Marks have the same meanings as in figure11.
\end{figure}
\begin{figure}
\includegraphics[width=10cm,clip]{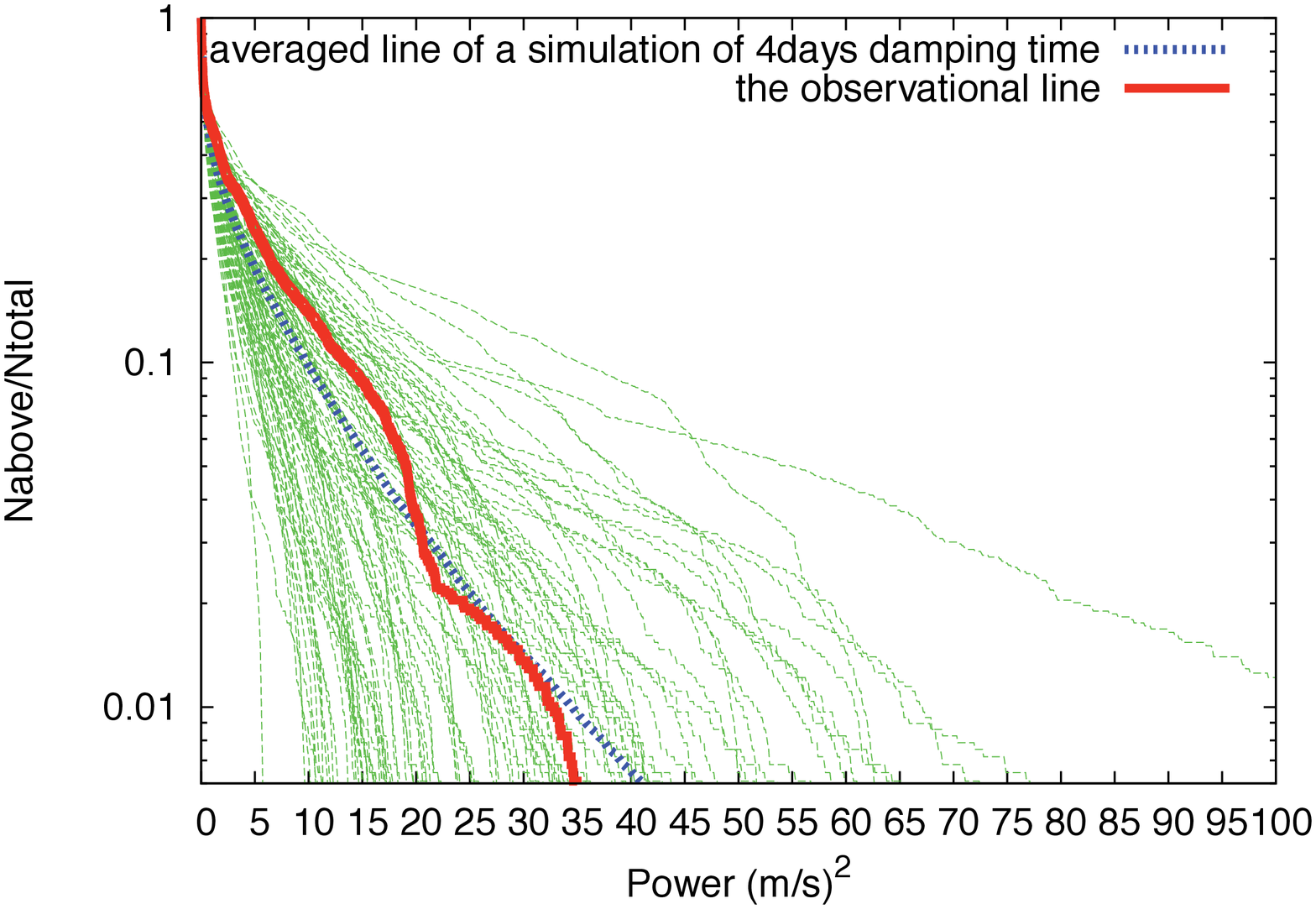} \\
 Fig. 15  Simulated cumulative power spectra of 100 models for $\zeta$ Hya (thin dotted line)\\
 \hspace{13mm} with damping time, 4 days,  and the observational cumulative power spectrum \\
\hspace{13mm} for data obtained in 2006 (thick solid line).  \\
 \hspace{13mm} Thick dotted line shows the averaged line of a simulation. \\
 \hspace{13mm} The ordinate shows fraction of the power above this level to the total power.  \\
 \hspace{13mm}
\end{figure}
\begin{figure}
\includegraphics[width=10cm,clip]{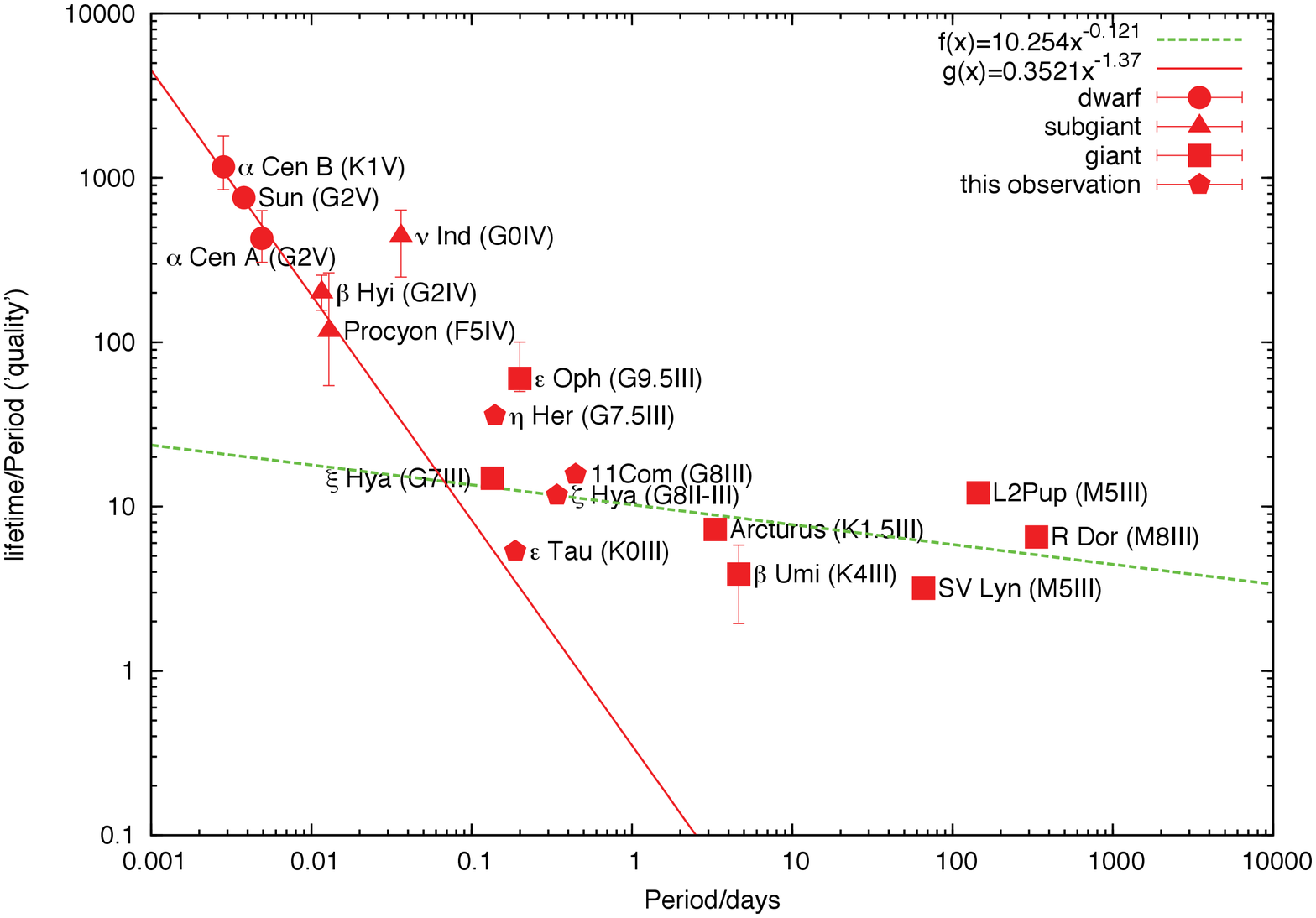} \\
 Fig. 16  Q-values of our 4 giants are plotted against their periods (filled pentagon)  \\
 \hspace{13mm} in the diagram created by \cite{S-4}(2006).  \\
 \hspace{13mm} Solid line shows best fit line for dwarf and subgiant groups except for $\nu$ Ind. \\
 \hspace{13mm} Dotted line shows best fit line for G-, K-, and M-giant groups. \\
 \hspace{13mm} $\alpha$ Cen A,B: \cite{K-3}(2005), Sun: \cite{C-4}(1997), \\
 \hspace{13mm} $\nu$ Ind: \cite{C-3}(2007), $\beta$ Hyi: \cite{B-4}(2007), \\
 \hspace{13mm} $\alpha$ CMi: \cite{A-2}(2008), $\epsilon$ Oph: \cite{K-1}(2008), \\
 \hspace{13mm} $\xi$ Hya: \cite{S-4}(2006), $\alpha$ Boo: \cite{T-2}(2007), \\
 \hspace{13mm} $\beta$ UMi: \cite{T-3}(2008), $\rm{L}_2$ Pup: \cite{B-3}(2005) \\  
 \hspace{13mm} SV Lyn and R Dor: \cite{D-4}(2004)  
\end{figure}

\end{document}